\documentclass[superscriptaddress,onecolumn,floats,10pt,floatfix,nobibnotes,notitlepage,showkeys,longbibliography]{revtex4-1}

\usepackage{amsmath}
\usepackage{amsfonts}
\usepackage{amssymb}

\usepackage{graphicx}
\usepackage{subfigure}

\usepackage[outdir=./]{epstopdf} 
\DeclareGraphicsExtensions{.eps}

\usepackage{color}
\usepackage[colorlinks=true,citecolor=blue,linkcolor=blue,urlcolor=black]{hyperref}
\usepackage{times}
\usepackage{amstext}
\usepackage{latexsym}
\usepackage{bbm}
\usepackage[normalem]{ulem}
\usepackage{braket}		
\usepackage[utf8]{inputenc}
\usepackage[english]{babel}
\usepackage[T1]{fontenc}
\usepackage{lmodern}
\usepackage{bbold} 		
\usepackage{amsmath} 	
\usepackage{amsfonts} 	
\usepackage{appendix}
\usepackage{stackengine}

\providecommand{\bra}[1]{\langle #1 \rvert}
\providecommand{\ket}[1]{\lvert #1 \rangle}
\providecommand{\be}{\begin{equation}}
\providecommand{\ee}{\end{equation}}
\providecommand{\ba}{\begin{eqnarray}}
\providecommand{\ea}{\end{eqnarray}}

\usepackage{mathbbol}
\usepackage{mathtools}

\begin{document}
\title{Work and Heat Value of Bound Entanglement}

\author{Asl\i \ Tuncer}
\email{asozdemir@ku.edu.tr}
\affiliation{Department of Physics, Koc University, 34450 Sariyer, Istanbul, Turkey}
\author{Mohsen Izadyari}
\email{mizadyari18@ku.edu.tr}
\affiliation{Department of Physics, Koc University, 34450 Sariyer, Istanbul, Turkey}
\author{Ceren B. Da\u{g}}
\email{cbdag@umich.edu}
\affiliation{Physics Department, University of Michigan, Ann Arbor, Michigan 48109, USA}
\author{Fatih Ozaydin}
\email{fatih.ozaydin@isikun.edu.tr}
\affiliation{ Institute for International Strategy, Tokyo International University, 1-13-1 Matoba-kita, Kawagoe, Saitama, 350-1197, Japan}
\affiliation{Department of Information Technologies, Isik University, Sile, Istanbul, 34980, Turkey}
\author{\"{O}zg\"{u}r E.~M\"{u}stecapl{\i}o\u{g}lu}
\email{omustecap@ku.edu.tr}
\affiliation{Department of Physics, Koc University, 34450 Sariyer, Istanbul, Turkey}


\begin{abstract}
Entanglement has recently been recognized as an energy
resource which can outperform classical resources if decoherence is relatively low.
Multi-atom entangled states can mutate irreversibly to so called bound entangled (BE) states under noise. Resource value of BE states in information applications
has been under critical study and a few cases where they can be useful have been identified.
We explore the energetic value of typical BE states. Maximal work extraction is determined in terms of ergotropy. Since the BE states are non-thermal, extracting heat from them is less obvious. We compare single and repeated interaction schemes to operationally define and harvest heat from BE states. BE and free entangled (FE) states are compared in terms of their ergotropy and maximal heat values.
Distinct roles of distillability in work and heat values of FE and BE states are pointed out. Decoherence effects in dynamics of ergotropy
and mutation of FE states into BE states are examined to clarify significance of the work value of BE states. Thermometry of distillability of entanglement using micromaser cavity is
proposed.
\end{abstract}

\keywords{Quantum Entanglement; Quantum Coherence; Quantum Thermodynamics}

\maketitle

\section{Introduction}
\label{intro}
Bound entanglement is a unique form of entanglement in the sense that it is irreversible: Entanglement is necessary to prepare it, however no entanglement can be distilled from it via local operations and classical communication (LOCC)~\cite{horodecki_mixed-state_1998,horodecki_quantum_2009}. A Bound entangled (BE) state is nondistillable,
having a positive partial transpose on the contrary to
free entangled (FE) states.
Irreversibility, which is fundamental for formation of all nondistillable entangled states~\cite{yang_irreversibility_2005},
has been discussed from the view point of thermodynamics as well~\cite{brandao_entanglement_2008,horodecki_are_2002,brandao_reversible_2010,horodecki_quantum_2008}.
Energy value of quantum states from resource theory and thermodynamic point of views have been
attracted much attention
recently~\cite{brandao_resource_2013,allahverdyan_maximal_2004,francica_daemonic_2017,fusco_work_2016,hsieh_work_2017,brandner_universal_2017}. In particular, quantum coherence and entanglement have been considered to power
up quantum heat engines which can outperform their classical counterparts~\cite{scully_extracting_2003,turkpence_quantum_2016,dag_multiatom_2016,niedenzu_operation_2016,niedenzu_quantum_2018}, being a practical motive in
the emerging field of
quantum thermodynamics~\cite{millen_perspective_2016,allahverdyan_extraction_2000,weimer_local_2008,mahler_quantum_2014,kosloff_quantum_2013,vinjanampathy_quantum_2016,goold_role_2016,hardal_phase-space_2018,uzdin_equivalence_2015,bauer_optimal_2016,brandner_periodic_2016}.
While single qubit quantum coherence is not sufficient to power up practical systems subject to decoherence~\cite{quan_quantum-classical_2006}, higher dimensional systems can overcome the decoherence challenge~\cite{turkpence_quantum_2016,dag_multiatom_2016,turkpence_photonic_2017}.

Bipartite multi-qubit entangled states are high-dimensional quantum coherent systems that can be either in BE or in FE classes. Some FE states
can irreversibly turn into BE states under local noise~\cite{song_sudden_2009}. Though BE states
can be generated experimentally~\cite{amselem_experimental_2009,lavoie_experimental_2010,kaneda_experimental_2012}, natural presence of BE states
under thermal noise in many-body systems suggests that it can be a natural quantum entanglement resource for quantum energy processing~\cite{horodecki_mixed-state_1998,ferraro_thermal_2008,toth_optimal_2007,cavalcanti_distillable_2008}.
Moreover, it can be used per se for such applications, in contrast to quantum information processing where
it requires an activation~\cite{horodecki_bound_1999,horodecki_secure_2005,smith_quantum_2008,czekaj_quantum_2015,toth_quantum_2018}. Hence, it is of practical as well as of fundamental interest to examine energetic resource values of FE and BE states relative to each other.

\begin{figure}[t]
\centering
\includegraphics[width=3.5 cm]{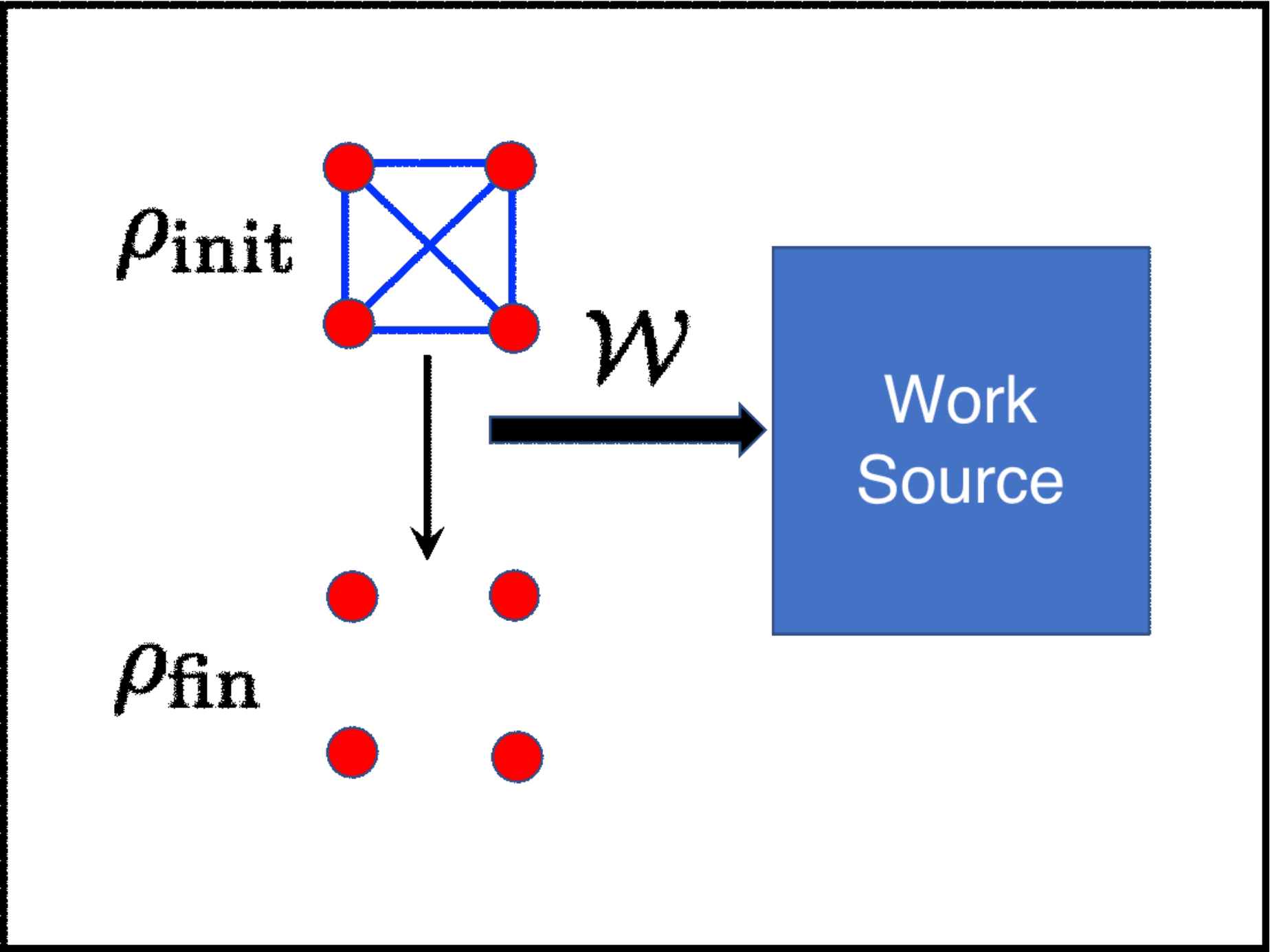}  \quad\quad
\includegraphics[width=5 cm]{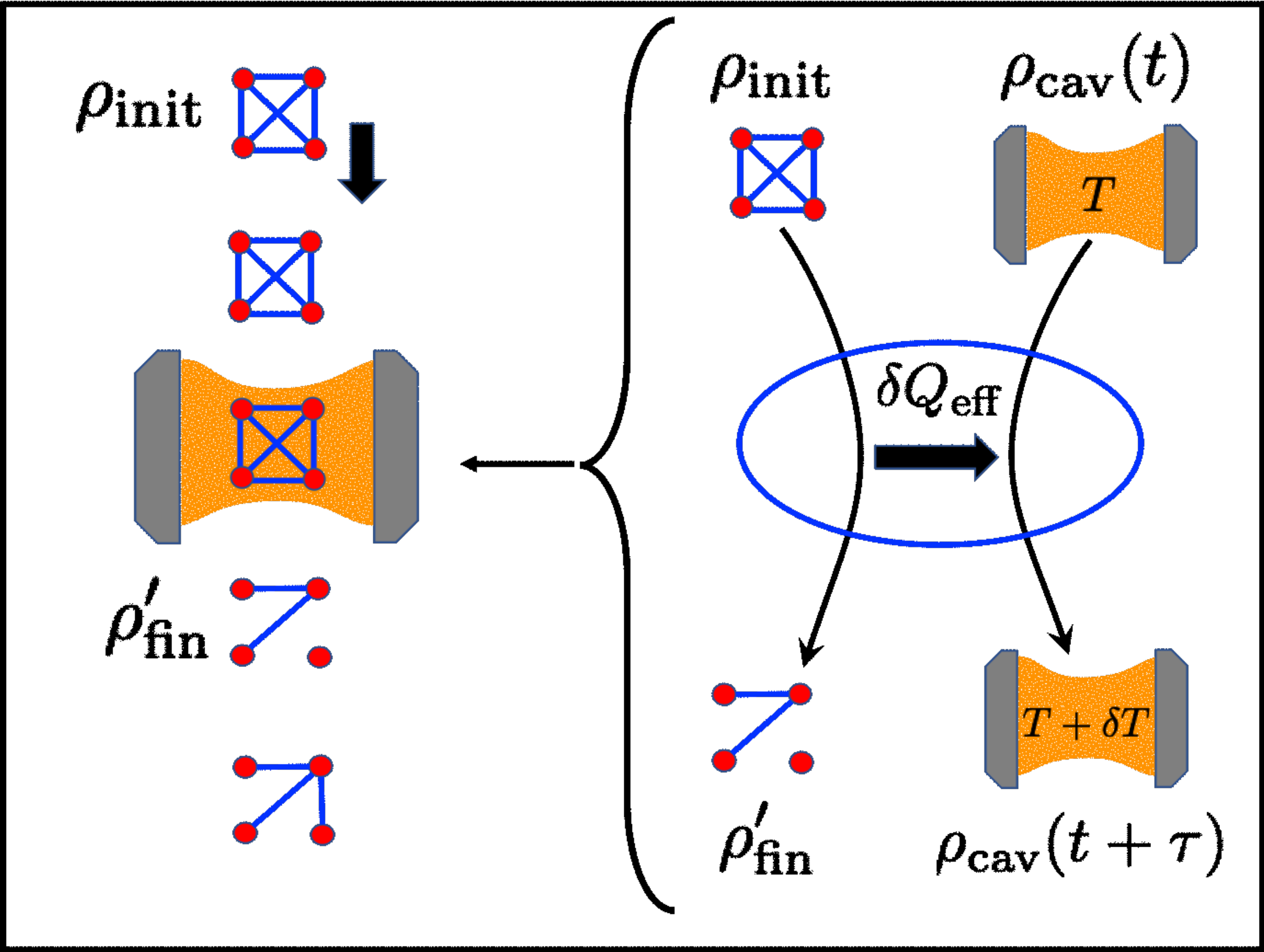}
\caption{(Left Panel) Maximum work extraction from four qubit bound entangled state $\rho_{\text{init}}$.
Red dots represent the qubits and the lines connecting them represent the coherences among them in the energy basis
of non-interacting qubits. The coherences are consumed after the interaction with the work source. If the interaction is optimal no coherence
is left in the final state $\rho_{\text{fin}}$ and the extracted work will be maximum, which is called as ``ergotropy'' denoted by ${\cal W}$.
(Right Panel) Effective heat extraction from four qubit bound entangled state $\rho_{\text{init}}$. Qubit clusters are injected repeatedly into a
resonator one at a time. Transition time of each cluster through the resonator is $\tau$. The interaction is not optimum and some coherences
can remain in the final state $\rho_{\text{fin}}^\prime$ after the transition. If the resonator field is initially a thermal equilibrium state
$\rho_{\text{cav}}(t)$ at temperature
T and if the interaction yields another Gibbsian state of the cavity $\rho_{\text{cav}}(t+\tau)$ at an emprical temperature $T+\delta T$
then the harvested coherent energy from the qubit cluster can be envisoned as effective
heat, $\delta Q_{\text{eff}}$, changing only disorder of the cavity field
without inducing any coherence. The bound entangled (BE) states considered in the text can perform this task. Their coherences cannot be translated
into the cavity field under Tavis-Cummings type interaction. Repeated interaction of the resonator with the clusters prepared in the same
BE state brings the cavity to a steady state at a certain temperature depending on the cluster state populations and coherences. Hence the beam of atomic clusters act as an effective heat bath with which the resonator can be brought into effective thermal equilibrium at
a genuine thermodynamic temperature. }
\label{fig:schemes}
\end{figure}

We specifically consider here typical four qubit~\cite{smolin_four-party_2001, fei_class_2006} as well as 
two qutrit BE states~\cite{horodecki_bound_1999}. 
We determine the maximal extractable work from these states by calculating their ergotropy~\cite{allahverdyan_maximal_2004} (cf.~left panel of Fig.~\ref{fig:schemes}). Besides, we examine dynamics of ergotropy when FE states change into BE states under amplitude damping~\cite{ali_distillability_2010}. Advantages of BE states as potentially natural and robust ergotropy resources are exemplified.
Complementary to the notion of single shot work extraction, here we explore the missing piece of the puzzle: we ask if and how we can make similar definitions
for heat extraction out of a general non-thermal quantum state using a single unitary operation. 
We follow an operational approach where we define work~\cite{gallego_thermodynamic_2016} and quantify heat by
the properties of an auxiliary system (cf.~right panel of Fig.~\ref{fig:schemes}). For that aim, a single mode cavity or a single qubit is introduced  as a "thermometer" system whose field remains in a canonical thermal (Gibbsian) distribution after a single interaction with
a multi-qubit cluster that is the "fuel" system. In addition to the single interaction route to heat harvesting, we consider a repeated interaction scheme. 
Following the concept of ergotropy, we determine maximal heat transfer (which may be called thermotropy) between the
thermometer system and an ensemble of identically prepared BE states after many interactions. The empirical temperature of the thermometer system
at the steady state is then used as a simple operational quantifier of thermotropy. We compare the achievable thermometer temperatures by single and repeated
interaction schemes and determine the optimum interaction time in the single interaction scheme that can yield higher temperatures than the repeated
interaction method. While operational definition of heat for a single interaction can be of fundamental interest, similar to single shot operational~\cite{gallego_thermodynamic_2016}  or
maximum extractable work definitions~\cite{allahverdyan_maximal_2004}, our results can have practical significance too by being simpler and faster
alternatives to early proposals of quantum thermalization by repeated interactions~\cite{scully_extracting_2003,turkpence_quantum_2016,dag_multiatom_2016,niedenzu_operation_2016,niedenzu_quantum_2018}.

This paper is organized as follows. In Sec.~\ref{sec:work} we present an ergotropic analysis of typical BE states~\cite{smolin_four-party_2001, fei_class_2006,horodecki_bound_1999}  in three subsections. In the fourth subsection, 
we examine the dynamics of ergotropy in an amplitude damping environment which can
mutate FE states into BE states.  Sec.~\ref{sec:heatDefinition} introduces an operational definition of heat, which is used in the
subsequent discussion of heat extraction from BE states in Sec.~\ref{sec:heat}. Two routes of heat harvesting, 
 single interaction or repeated interactions, are investigated and compared with each other in subsections 
 Sec.~\ref{sec:SS} and Sec.~\ref{sec:micromaser}, respectively. Summary of our results and our 
conclusion are given in Sec.~\ref{sec:con}.

\section{Maximum Work Extraction from Bound Entangled States}
\label{sec:work}

\subsection{Smolin Bound Entangled State}
\label{sec:smolinWork}
We consider a thermally isolated system of four non-interacting qubits, with the same transition frequency $\omega$, described by a Hamiltonian (we take $\hbar=1$)
\begin{equation}
H=\frac{\omega}{2}\left(\sigma_3^\mathrm{A}+\sigma_3^\mathrm{B}+
\sigma_3^\mathrm{C}+\sigma_3^\mathrm{D}\right).
\label{eq:H_4qubit_1}
\end{equation}
The system is initially prepared in a BE state, introduced by Smolin~\cite{smolin_four-party_2001}.   This state can be expressed in the form~\cite{amselem_experimental_2009}
\begin{equation}
\rho_{\mathrm{S}}=\frac{1}{16}\sum_{i\in{0,1,2,3}}
\sigma_i^\mathrm{A}\otimes\sigma_i^\mathrm{B}\otimes\sigma_i^\mathrm{C}\otimes\sigma_i^\mathrm{D},
\label{eq:Smolinstate}
\end{equation}
where $A,B,C,D$ label the four qubits. The components of the
Pauli spin matrices are denoted by $\sigma_i$ with $i=1,2,3$ and $\sigma_0=\mathbb{1}$ stands for the unit matrix.

Smolin state is a special four-qubit BE state with remarkable symmetry properties under exchange of qubits. It is BE in the sense that it is separable, and hence non-distillable, for every two-qubit partitions (e.g. AB|CD) but it is entangled for every single-party partitions (e.g. A|BCD). In contrast to other typical BE states however, it does not show positive partial transpose and non-separability properties simultaneously. Accordingly, bound entanglement in the Smolin state is ``unlockable" so that maximal entanglement can be distilled out of it by bring any two qubit parties together.~\cite{smolin_four-party_2001, BESM1, BES2, horodecki_mixed-state_1998, czekaj_quantum_2015} Even though it is not a FE but BE state, it can still make a useful quantum resource for energy harvesting as we will examine below.

\begin{figure}[!t]
	\centering
	\includegraphics[width=8.5cm]{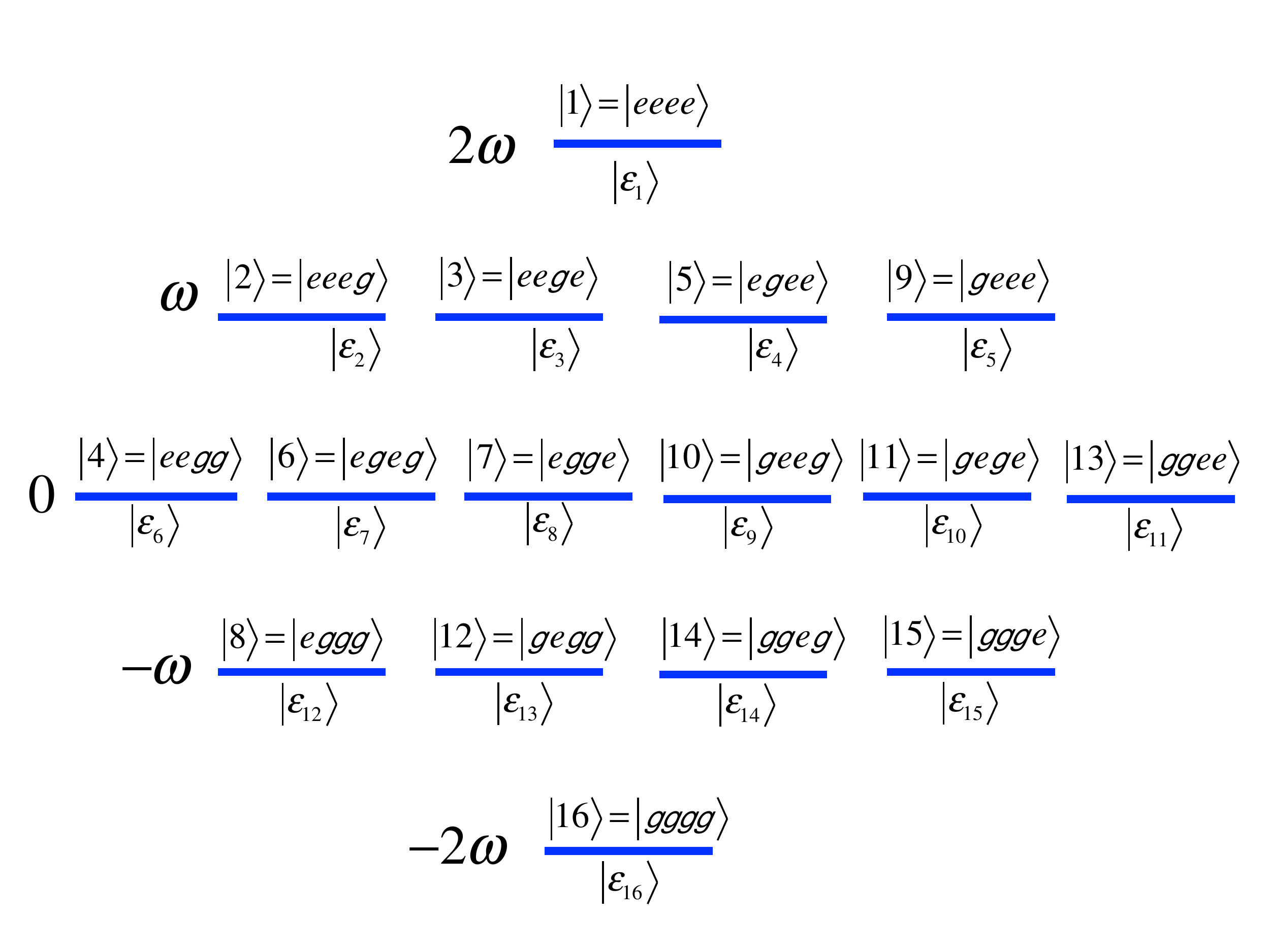}
	\caption{Relation between the computation basis and the energy basis for a four qubit system. Four qubit computational basis is given by
		$|abcd\rangle=|n\rangle$ with $a,b,c,d\in {0,1}$ where $n=1+8a+4b+2c+d$. The energy basis is given
		in terms of the excited ($e$) and ground ($g$) qubit states such that $x_1,x_2,x_3,x_4\in\{e,g\}$; they are numbered as $ |\epsilon_k\rangle$
		with $k=1..16$. For single qubit, the energy basis is related to the computational basis by $|e\rangle=|0\rangle=(1\; 0)^T$ and
		$|g\rangle=|1\rangle=(0\; 1)^T$.}
	\label{fig:bases}
\end{figure}

In order to harvest work from the system, it is coupled to an external work source such that a cyclic work transfer interaction $V(t)$ is applied to it from time $t=0$ to $t=\tau$ (cf.~the left panel in Fig.~\ref{fig:schemes}). Initially
the system has $E_0=0$ energy. If we can find the system at a minimum final energy after the cyclic process and if the final energy
is negative then the system does maximal work on the external work source. This problem has been solved
for general situations and the concept of ergotropy has been introduced corresponding to maximum work extraction from finite quantum
systems~\cite{allahverdyan_maximal_2004}. Ergotropy is expressed as
\begin{equation}
{\cal W}=\sum_{j,k}r_j\epsilon_k\left(|\langle r_j|\epsilon_k\rangle|^2-\delta_{jk}\right).
\label{eq:W}
\end{equation}
Here $r_j=1/4$ for $j=1..4$ and $r_j=0$ for $j=5..16$ are the eigenvalues of $\rho_{\mathrm{S}}$ in descending order and corresponding eigenvectors of nonzero eigenvalues are given in the appendix~\ref{sec:app0}. Eigenvalues of $H$ are denoted by $\epsilon_k$ and listed in ascending order, $\epsilon_k=\{-2,-1,..,0,..,1,..,2\}\omega$ with $k=1..16$ and corresponding eigenvectors, $ |\epsilon_k\rangle$, are given in Fig.~\ref{fig:bases}. The degeneracy factors of the distinct eigenvalues $\epsilon_k$ are $1,4,6,4,1$. The final energy of the system
\begin{equation}\label{eq:finalEnSmolin}
E_f=\sum_{j=1}^{16}r_j\epsilon_j=-1.25\omega
\end{equation}
is the minimum energy and the maximum extractable work is found to be ${\cal W}=1.25\omega$. 

After the work extraction the Smolin state is transformed into
\begin{eqnarray}\label{eq:finalRho}
\rho_f&=&\sum_jr_j|\epsilon_j\rangle\langle\epsilon_j|,\\
&=&\frac{1}{4}\left(|gggg\rangle\langle gggg|+|ggge\rangle\langle ggge|+|ggeg\rangle\langle ggeg|+|gegg\rangle\langle gegg|\right).
\end{eqnarray}
We express the states in the energy basis of the non-interacting four qubit system; excited and ground states of the qubits are denoted
by $|e\rangle,|g\rangle$ which correspond to the computational basis states $|0\rangle,|1\rangle$. Four qubit computational basis is given by
$|abcd\rangle=|n\rangle$ with $a,b,c,d\in {0,1}$ where $n=1+8a+4b+2c+d$. The relation between the energy and computation bases
is given in Fig.~\ref{fig:bases}. The eigenvalues of $\rho_{\mathrm{S}}$ are preserved under unitary work extraction. 
Optimum
final state distributes the larger eigenvalues to the lower energy levels. $\rho_{\mathrm{S}}$ is not diagonal in the energy basis; it
has coherences which are harvested as work under cyclic application of $V(t)$.
If $V(t)$ is optimal then all the energy coherences are completely harvested such that
the final state is of the form Eq.~(\ref{eq:finalRho}), which is diagonal in the energy basis (cf. left panel of Fig.~\ref{fig:schemes}).  Let us note that $\rho_f$ is not symmetric under qubit exchange. However qubit exchanges will only swap corresponding eigenvectors $|r_j\rangle$ between two manifolds associated with eigenvalues $0$ and $1/4$; accordingly the structure of the expression Eq.~(\ref{eq:finalRho}) remains the same with the summation index counting the states of non-zero eigenvalue. Hence, the final energy in Eq.~(\ref{eq:finalEnSmolin}) is invariant under qubit exchange (cf.~Fig.~\ref{fig:bases} where $\epsilon_j$ levels can be seen invariant under qubit exchanges).

Optimum $V(t)$ can be determined
in principle from the map~\cite{allahverdyan_maximal_2004} $\rho_S\mapsto\rho_{\mathrm{f}}=U\rho_{\mathrm{S}} U^\dagger$,
where the optimum unitary $U$ is given by
\begin{equation}
U=\sum_{j=1}^{16}|\epsilon_j\rangle\langle r_j|.
\end{equation}
 We remark that to calculate the ergotropy the explicit forms of $U$ or $V(t)$ are not required. As $U$ is completely determined by the eigenvectors of both the reference hamiltonian $H$ and the initial state $\rho_{\mathrm{S}}$, it differs for different choices of $H$. On the other hand, contribution of $U$ to the ergotropy, through $\rho_f$ in ${\cal W}=\text{tr}(\rho_{\mathrm{S}}H)-\text{tr}(\rho_{\mathrm{f}}H)$, only depends on the eigenspectrum of $H$ and $\rho_{\mathrm{S}}$. Accordingly, that contribution is invariant under unitary basis changes or simple rotations. Hence, up to reference energy, ergotropy captures the basis independent maximal extractable energy, or non-passivity, of a given quantum state. In our calculations, $H$ is taken to be that of non-interacting qubits (or qutrits). Such a choice allows for assessing the ergotropic content of the initial bound entangled states per se, by avoiding potential ambiguities that might arise from contributions of interactions or coherences induced by the reference Hamiltonian.

Despite being diagonal and commutative with $H$, the final state is not a Gibbsian. Intuitively the optimum work
extraction could be possible if the Smolin state maps to a thermal equilibrium state at zero temperature effectively the ground state, for which $E_f=-2\omega$.
For finite systems subject to unitary work extraction, ergotropy is, in general, less than or equal to the case where the final state is a Gibbsian, such
that we have ${\cal W}\le {\cal W}_{\mathrm{th}}$~\cite{allahverdyan_maximal_2004}. Smolin state cannot yield optimum ergotropy ${\cal W}_{\mathrm{th}}$.

Let us now compare performance of Smolin state with both separable states and with FE states. In fact we can make more general statements
for all BE states using a simple deductions. It is proven in Ref.~\cite{allahverdyan_maximal_2004} that if a state $\rho$ majorizes~\cite{Bhatia_matrix_1997} another state $\sigma$ ($\rho\succ\sigma$) with the
same energy, then $\rho$ has higher ergotropy. As pure states majorizes all the other
states, one can always find either a separable state or a maximally entangled state that
would majorize a BE state, which is not pure. On the other hand, while maximally entangled state or pure separable states are always more active than BE states, there can
be non-maximal FE states which perform poorly relative to BE states at the same energy.
In the subsequent discussion, we shall consider a parametrized class of bound entangled state for further comparison of the work harvesting from BE states
relative to FE states.
\subsection{A Class of Bound Entangled States: FLS state}
\label{sec:classWork}
We will now consider a class of bound entangled states, which we dub as FLS state, as it is introduced in Ref.~\cite{fei_class_2006} by S.~M.~Fei, X.~Li-Jost, 
and B.~Z.~Sun. It is parametrized
with a parameter $\varepsilon$ and expressed in the computational basis as

\begin{equation}
\rho_{\varepsilon}=
\left(
  \begin{array}{cccccccccccccccc}
    {1- \varepsilon \over 4} & 0 & 0 & 0 & 0 & 0 & 0 & 0 & 0 & 0 & 0 & 0 & 0 & 0 & 0 & 0 \\
    0 &  {\varepsilon \over 8} & 0 & 0 & -{\varepsilon \over 8} & 0 & 0 & 0 & 0 & 0 & 0 & 0 & 0 & 0 & 0 & 0 \\
    0 & 0 & {\varepsilon \over 8} & 0 & 0 & 0 & 0 & 0 & -{\varepsilon \over 8} & 0 & 0 & 0 & 0 & 0 & 0 & 0 \\
    0 & 0 & 0 & 0 & 0 & 0 & 0 & 0 & 0 & 0 & 0 & 0 & 0 & 0 & 0 & 0 \\
    0 & - {\varepsilon \over 8} & 0 & 0 & {\varepsilon \over 8} & 0 & 0 & 0 & 0 & 0 & 0 & 0 & 0 & 0 & 0 & 0 \\
    0 & 0 & 0 & 0 & 0 & {1- \varepsilon \over 4} & 0 & 0 & 0 & 0 & 0 & 0 & 0 & 0 & 0 & 0 \\
    0 & 0 & 0 & 0 & 0 & 0 & 0 & 0 & 0 & 0 & 0 & 0 & 0 & 0 & 0 & 0 \\
    0 & 0 & 0 & 0 & 0 & 0 & 0 & {\varepsilon \over 8} & 0 & 0 & 0 & 0 & 0 & -{\varepsilon \over 8} & 0 & 0 \\
    0 & 0 & -{\varepsilon \over 8} & 0 & 0 & 0 & 0 & 0 & {\varepsilon \over 8} & 0 & 0 & 0 & 0 & 0 & 0 & 0 \\
    0 & 0 & 0 & 0 & 0 & 0 & 0 & 0 & 0 & 0 & 0 & 0 & 0 & 0 & 0 & 0 \\
    0 & 0 & 0 & 0 & 0 & 0 & 0 & 0 & 0 & 0 & {1- \varepsilon \over 4} & 0 & 0 & 0 & 0 & 0 \\
    0 & 0 & 0 & 0 & 0 & 0 & 0 & 0 & 0 & 0 & 0 & {\varepsilon \over 8} & 0 & 0 & -{\varepsilon \over 8} & 0 \\
    0 & 0 & 0 & 0 & 0 & 0 & 0 & 0 & 0 & 0 & 0 & 0 & 0 & 0 & 0 & 0 \\
    0 & 0 & 0 & 0 & 0 & 0 & 0 & -{\varepsilon \over 8} & 0 & 0 & 0 & 0 & 0 & {\varepsilon \over 8} & 0 & 0 \\
    0 & 0 & 0 & 0 & 0 & 0 & 0 & 0 & 0 & 0 & 0 & -{\varepsilon \over 8} & 0 & 0 & {\varepsilon \over 8} & 0 \\
    0 & 0 & 0 & 0 & 0 & 0 & 0 & 0 & 0 & 0 & 0 & 0 & 0 & 0 & 0 &  {1- \varepsilon \over 4} \\
  \end{array}
\right).
\label{eq:BES_matrix}
\end{equation}

It is bound entangled for $0 \leq \varepsilon \leq 0.5$ and free entangled for $0.5 < \varepsilon \leq 1$. This condition depends
on partitioning of the Hilbert spaces on which $\rho_{\varepsilon}$ is constructed~\cite{cheng_comment_2007}.
Here we assume $\rho_{\varepsilon}$ is constructed over
a tensor product of Hilbert spaces of dimensions $4\times 4$~\cite{fei_class_2006}. A more general, $5$ parameter
state with the same structure of $\rho_{\varepsilon}$ is introduced in Ref.~\cite{fei_class_2006}; it is shown
to be bound entangled under the same condition of $0 \leq \varepsilon \leq 0.5$,
which depends only the parameter $\varepsilon$~\cite{cheng_comment_2007}. Our objective here is to reveal qualitative
differences in the behavior of ergotropy of $\rho_{\varepsilon}$ with $\varepsilon$,  rather than optimization of ergotropy, which
would be possible by considering the more general parametrization of $\rho_{\varepsilon}$.

We use the same Hamiltonian $H$ given in Eq.~(\ref{eq:H_4qubit_1}), taking the same transition frequency $\omega$
for each qubit. Ergotropy can be obtained by using the eigenvalues and eigenvectors of $H$ and $\rho_{\varepsilon}$
in Eq.~(\ref{eq:W}). We need to distinguish two ranges of $\varepsilon$ when we list the eigenvalues of  $\rho_{\varepsilon}$
in descending order. For $0\le\varepsilon\le 0.5$, the eigenvalues of $\rho_{\varepsilon}$ in
descending order are
\begin{equation}
  r_j =
  \begin{cases}
     (1-\varepsilon)/4& \text{for}\; j=1 \dots 4; \\
    \varepsilon/4 & \text{for}\; j=5 \dots 8; \\
    0 & \text{for}\;  j=9 \dots 16.
  \end{cases}
\end{equation}
While for $0.5\le\varepsilon\le 1$, the eigenvalues of of $\rho_{\varepsilon}$ in descending order are
\begin{equation}
  r_j =
  \begin{cases}
    \varepsilon/4 & \text{for}\; j=1 \dots 4; \\
    (1-\varepsilon)/4 & \text{for}\; j=5 \dots 8; \\
    0 & \text{for}\;  j=9 \dots 16.
  \end{cases}
\end{equation}
The eigenvalues of the Hamiltonian $H$ in ascending order is given in the previous section. Calculating the ergotropy we
find
\begin{equation}
  {\cal W} =\omega
  \begin{cases}
    1.25-\varepsilon & \text{for}\;0\le\varepsilon\le 0.5; \\
    0.25+\varepsilon & \text{for}\;0.5\le\varepsilon\le 1.
  \end{cases}
  \label{eq:ergotropyFei}
\end{equation}

Initial energy of $\rho_{\varepsilon}$ is independent of $\varepsilon$, $E_0=0$, hence we can
make a meaningful comparison of ergotropies as resource values of states
at different $\varepsilon$.
In the bound entanglement domain,
ergotropy of $\rho_{\varepsilon}$ linearly decreases with $\varepsilon$, while in the free entanglement domain, it is
linearly increasing with $\varepsilon$. We see that for a given BE state $\rho_\epsilon$ with
$\varepsilon\le 0.5$, one can find a set of FE states $\rho_\alpha$ with
$\alpha\le 1-\varepsilon$ which
perform poorly relative to the BE state such that  ${\cal W}(\rho_\varepsilon)\ge{\cal W}(\rho_\alpha)$.
Equality happens at $\varepsilon=0$. Maximum ergotropy that is achievable by this
parametrized family of BE states at $\varepsilon=0$ and $\varepsilon=1$ is equal to that of Smolin state.

We can comment on this behavior from a constructing expression
of $\rho_{\varepsilon}$ in the form~\cite{fei_class_2006}
\begin{eqnarray}\label{eq:rhoeps}
\rho_{\varepsilon}=(1-\varepsilon){\cal I}_4+\varepsilon\rho_{abcd}.
\end{eqnarray}
Here $\rho_{abcd}$ is a free entangled state with negative partial transpose while ${\cal I}_4$ is a BE state with
positive partial transpose. $ {\cal W}({\cal I}_4)={\cal W}(\rho_{abcd})$ indicates that the
entangled states ${\cal I}_4$ and $\rho_{abcd}$ have the same resource values regardless of their opposite distillability character.
If we mix them however, the resultant state has less resource value than the components.
The state with maximally mixed distillability and nondistillability at $\varepsilon=0.5$
is the state with smallest ergotropy. Inseparability and nondistillability have distinct effects on the work resource value of $\rho_{\varepsilon}$. Symmetric behavior of ${\cal W}$ about $\varepsilon=0.5$ suggests that inseparability has the same positive effect for both BE and FE states; while
nondistillability makes positive and negative contribution to work values of BE and FE states, respectively.
The class of BE states $\rho_{\varepsilon}$ constitutes an example where a subset of FE states can always be found for a given BE state which can be
at least as valuable as FE states.
We can also compare Smolin state and $\rho_{\varepsilon}$ as they have the same initial energy. For any $\varepsilon$, ergotropy of
$\rho_{\varepsilon}$ is bounded from above by that of the Smolin state.
\subsection{Horodecki BE State of Two Qutrits }
\label{sec:HTQ}
Another commonly considered BE state is the so-called Horodecki state~\cite{horodecki_bound_1999} given by
\begin{equation}
\rho_\alpha=\frac{2}{7}\ket{\psi_+}\bra{\psi_+}+\frac{\alpha}{7}\sigma_++\frac{5-\alpha}{7}\sigma_-,
\label{eq:rho_alpha}
\end{equation}
where 
\begin{eqnarray}
|\psi_+\rangle&=&\frac{1}{\sqrt{3}}(|ge\rangle+|eg\rangle+|uu\rangle),\\
\sigma_+&=&\frac{1}{3}(|gg\rangle\langle gg|+|eu\rangle\langle eu|+|ge\rangle\langle ue|), \\
\sigma_-&=&\frac{1}{3}(|ee\rangle\langle ee |+|ug\rangle\langle ug| + |gu\rangle\langle gu|).
\end{eqnarray}
Here the energy levels are labeled by $g,u$ and $e$, respectively. This state
is separable for $2\le \alpha \le 3$, BE for $3< \alpha\le 4$, and FE for $4< \alpha \le 5$.
It can be written in the computational basis as
\begin{equation}
\rho_{\alpha}=
\left(
\begin{array}{ccccccccc}
{2\over21} & 0 & 0 & 0 & {2\over21}& 0 & 0 & 0 & {2\over21} \\
0 &  {\alpha\over 21} & 0 & 0 & 0 & 0 & 0 & 0 & 0\\
0 & 0 & {(5-\alpha) \over 21} & 0 & 0 & 0 & 0 & 0 & 0\\
0 & 0 & 0 & {(5-\alpha) \over 21}  & 0 & 0 & 0 & 0 & 0 \\
{2\over21} &0& 0 & 0 & {2\over21} & 0 & 0 & 0 & {2\over21} \\
0 & 0 & 0 & 0 & 0 &{\alpha\over 21} & 0 & 0 & 0 \\
0 & 0 & 0 & 0 & 0 & 0 & {\alpha\over 21} & 0 & 0  \\
0 & 0 & 0 & 0 & 0 & 0 & 0 &{(5-\alpha) \over21} & 0 \\
{2\over21} & 0 & 0& 0 & {2\over21} & 0 & 0 & 0 & {2\over21}\\
\end{array}
\right).
\end{equation}

\begin{figure}[t!]
	\centering
	\includegraphics[scale=0.35]{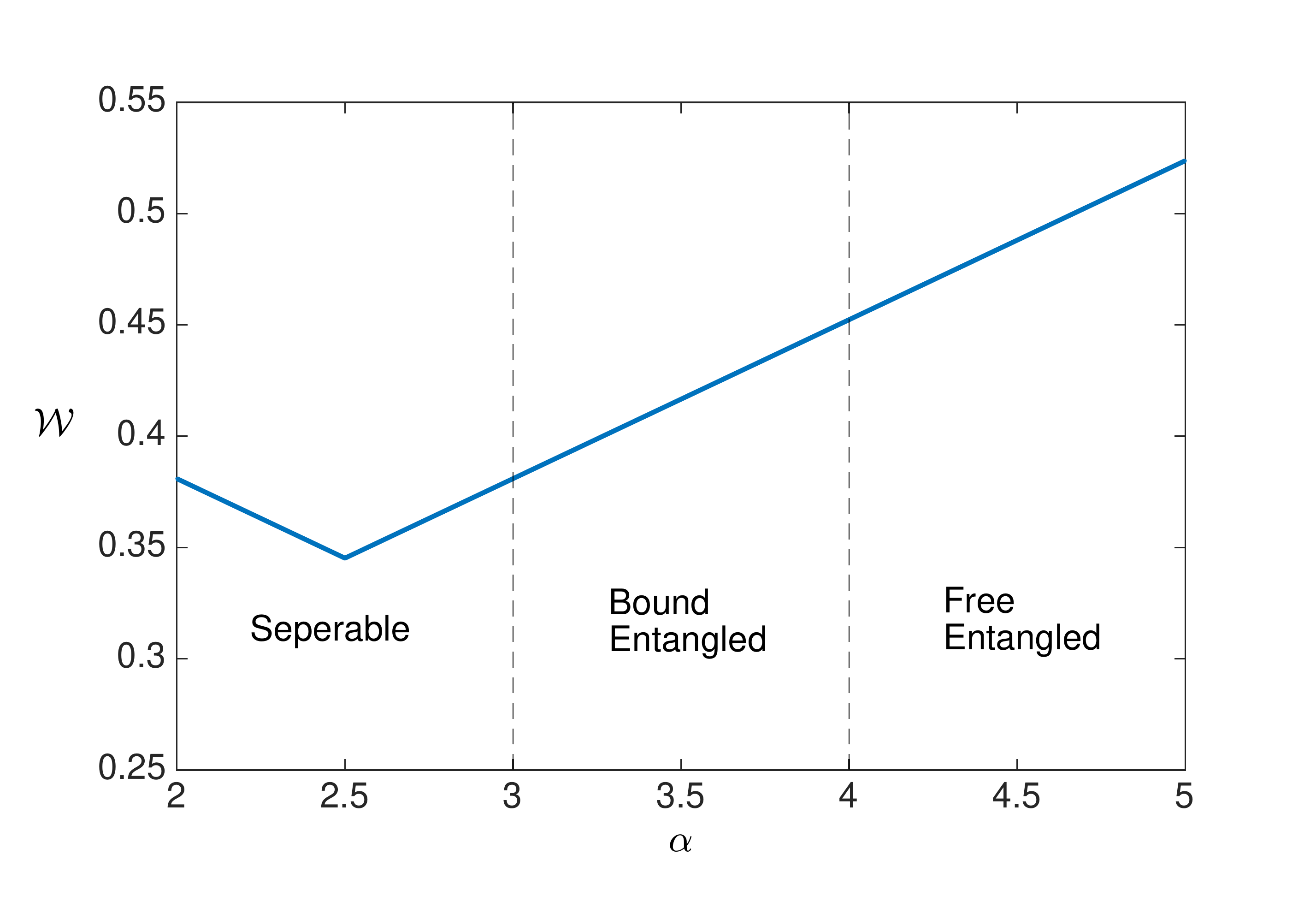}
	\caption{Ergotropy of the system of two 3-level atoms with respect to the parameter $\alpha$ determining whether the system is separable, bound entangled or free entangled. Ergotropy is linearly decreasing for seperable state in $2\le\alpha\le2.5$ region, and it is linearly increasing for $2.5\le\alpha\le5$ parameter values. Ergotropy of bound entangled state can be obtained as greater than the seperable state.} 
	\label{fig:2qutrit_BE}
\end{figure}

Assuming that the transition frequency $\omega$ is same for each qutrit, we consider the Hamiltonian for the system as
\begin{equation}
H=\frac{\omega}{2}\left(\sigma_3^\mathrm{A}+\sigma_3^\mathrm{B}\right),
\label{eq:H_2qutrits}
\end{equation}
where $\sigma_3$ is the generalized Pauli spin operator represented by a diagonal matrix whose elements are $1,0,-1$ in the basis of 
 $|e\rangle=(1 0 0)^\dagger$, $|u\rangle=(0 1 0)^\dagger$, and $|g\rangle=(0 0 1)^\dagger$. We consider V-type transition scheme for each qutrit of eigenenergies $E_e=\omega/2$, $E_u=\omega/2$, $E_g=-\omega/2$ for the corresponding eigenstates of $|e\rangle$, $|u\rangle$ and  $|g\rangle$, respectively. Ergotropy can be calculated via substituting the eigenvalues and eigenvectors of $H$ and $\rho_{\alpha}$ into Eq.~(\ref{eq:W}),
 which yields the ergotropy of $\rho_{\alpha}$ as
\begin{equation}
  {\cal W} =\omega
  \begin{cases}
    0.52144-0.071429\alpha & \text{for}\;2\le\alpha\le 2.5; \\
    0.16667+0.071429\alpha & \text{for}\;2.5\le\alpha\le 5,
  \end{cases}
  \label{eq:ergotropyHor}
\end{equation}
which is plotted in Fig.~\ref{fig:2qutrit_BE}. We see that ergotropy of the BE Horodecki state is lower than the FE and 
higher than the separable versions. Such a hierarchy does not exist for the FLS BE state for which BE and FE states exhibit
symmetric ergotropic values (cf.~Eq.~\ref{eq:ergotropyFei}). 
When we compare the ergotropies of two qutrits in Horedecki BE state and the four qubit Smolin and FLS BE states it can be seen easily that the ergotropy of two qutrits in Horedecki BE and FE states are smaller than the four qubit Smolin and both FLS BE and FE states.  

\subsection{Dynamical Mutation of Free to Bound Entanglement and Ergotropy Dynamics }
\label{sec:DFB}

\begin{figure}[!t]
	\centering
	\includegraphics[scale=0.4]{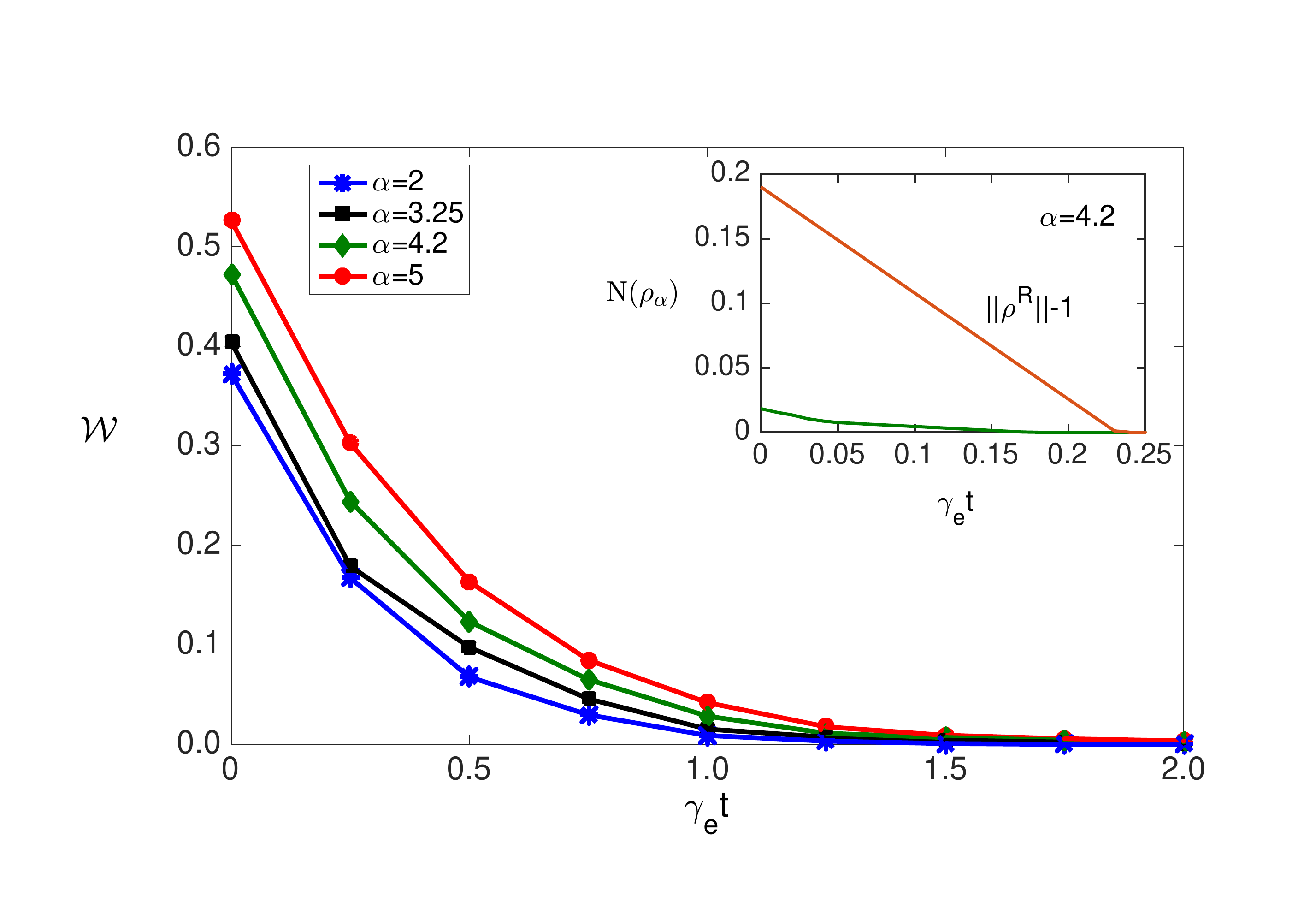}
	\caption{ Time evolution of ergotropy of two-qutrits initially in Horodecki state under local amplitude damping with $\gamma_u=\gamma_e/2$.
	The initial state is seperable for $\alpha=2$ (blue stars), bound entangled (BE) for $\alpha=3.25$ (black squares), and free entangled (FE) for $\alpha=4.2$ (green diamonds). 
	In the inset, the 
	negativity of the Horodecki state with $\alpha=4.2$ is shown~\cite{mazhar_ali}, which reveals that the initial FE state
	changes to a BE state at $\gamma_e t\approx0.1826$. It is found in Ref.~\cite{mazhar_ali} that the realignment parameter~\cite{chen_matrix_2003}, $||\rho^R(t)||-1$, is positive for $\alpha=4.2$ in the range $0.1826\leq\gamma_e t\leq0.2426$ so that the state is BE in this interval. Here we conclude that even if the state changes from FE to BE, it can still be a significant ergotropy resource for a range of $\gamma_et$.}
	\label{fig:2qutrit_BE_Dyn}
\end{figure}
%

Ideally one would prepare a cluster of atoms in FE states and then transfer them to a target system, such as an
optical cavity or a two-level atom, to harvest work. However, the atoms may be subject to open system 
decoherence during their transfer and hence
can lose their entanglement resource value partially or completely. An intriguing scenario is the mutation of FE states to BE states. An example is
given in Ref.~\cite{ali_distillability_2010} where two qutrits, subject to amplitude damping, exhibit 
distillability sudden death. Our objective is to examine the ergotropic value of the initial FE state during its mutation to BE state. For that aim, we consider the same
system as in Ref.~\cite{ali_distillability_2010} where two-qutrits in $V$-type transition scheme
decay to their own local reservoirs under amplitude damping as described by a master equation 
\begin{equation}
\label{eq:2Qt_ME}
\frac{d\rho}{dt}=\Gamma\rho,
\end{equation}
where
\begin{equation}
\label{ME_gama}
\Gamma\rho=\frac{\gamma_e}{2}\sum_{i=1,2}(2\sigma_{ge}^{(i)}\rho\sigma_{eg}^{(i)}-\sigma_{ee}^{(i)}\rho-\rho\sigma_{ee}^{(i)})
+\frac{\gamma_u}{2}\sum_{i=1,2}(2\sigma_{gu}^{(i)}\rho\sigma_{ug}^{(i)}-\sigma_{uu}^{(i)}\rho-\rho\sigma_{uu}^{(i)}).
\end{equation}
Here, the damping parameters are denoted by $\gamma_e,\gamma_u$. $\sigma_{kl}^{(1)}=\sigma_{kl}\otimes \mathbb{1}_3$ and $\sigma_{kl}^{(2)}=\mathbb{1}_3 \otimes \sigma_{kl}$ for $k,l=e,u,g$
are the transition operators for qutrits labelled by $(1)$ and $(2)$, respectively. The most general solution of the master equation~(\ref{eq:2Qt_ME}) has been given for an arbitrary initial density matrix  in~Ref.~\cite{ali_distillability_2010}.

The mutation from distillable to non-distillable entanglement or distillability  sudden death can be seen in the inset at the corner of Fig.~\ref{fig:2qutrit_BE_Dyn}, showing typical dynamical behavior of the negativity $N(\rho)$ of an 
initial FE state. Corresponding evolution of the ergotropy is plotted in the main figure for a set of $\alpha$ parameters. The ergotropies of FE states with different $\alpha$ parameters do not show a cross over behaviour in short time and hence the ergotropy can be thought as an identifier of the 
state at all times. When the state is no longer FE but BE, it can still possess
significant ergotropy. 

\section{An operational definition of heat quantifier }
\label{sec:heatDefinition}

Before presenting a specific discussion of heat value of bound entangled states, we would like to clarify the
notion of heat from a quantum state. Following the operational definition of work quantifiers~\cite{gallego_thermodynamic_2016},
which defines work by examining its effect on a target system, we put forward
use an operational definition of heat. We consider a system $S_Q$
for heat transfer in addition to the main system that we call as resource ($R$) $S_{R}$.
Let a global unitary $U_{RQ}$
act on the composite system $S_R\otimes S_Q$. We can determine the change
in the local (reduced) state $\rho_Q$ by a map
\begin{eqnarray}
\rho_Q\mapsto \rho_Q^\prime= \text{Tr}_R\left( U_{RQ}
\rho_R\otimes \rho_Q U_{RQ}^\dag\right),
\label{eq:thermalOp}
\end{eqnarray}
where the reduced states of $S_Q$ and $S_{R}$ are given by
$\rho_Q=\text{Tr}_R\rho_{RQ}$ and $\rho_R=\text{Tr}_Q\rho_{RQ}$.
The trace operator brings irreversibility to the operation. We define the heat based on the properties of the
state $\rho_Q^\prime$. A natural condition to request is that $\rho_Q^\prime$ is a Gibbsian object, which  is a classical-like state
such that it is diagonal in the energy basis with eigenvalues decreasing with energy. Moreover an empirical temperature
can be assigned to it when written in a Gibbsian form. Under these conditions,
we can identify the associated energy
change as a heat quantifier $\delta Q = \text{Tr}(H_Q\rho_Q^\prime)- \text{Tr}(H_Q\rho_Q)$, where
$H_Q$ is the Hamiltonian of $S_Q$. Then the matrix form of the heat quantifier ${\cal \delta Q}:=\text{max}({\delta Q})$ could be called ``thermotropy'', as the heat analog of the term ergotropy.

In this paper, we restrict ourselves to the systems $S_Q$ which are initially in thermal equilibrium states, for simplicity as well as for its relevance to both practical applications and to fundamental resource theories~\cite{goold_role_2016}. However, the operational definition of the quantifier for the effective heat can be extended to include initially coherent states for the target system as long as the
coherence in these states vanish in time so that the final state of the target system can be identified as a Gibbsian.
As both $\rho_Q$
and $\rho_Q^\prime$ are Gibbsian objects, the transformation becomes a generalized thermal operation~\cite{janzing_thermodynamic_2000,horodecki_fundamental_2013} which may be called as generalized
Gibbs-preserving map (GGPM)~\cite{janzing_thermodynamic_2000,faist_gibbs-preserving_2015,wilming_second_2016}.
We use the term ``generalized'' to
distinguish our case from the usual definitions which require the Gibbsians with the same energy or temperature before and after the transformation.
While the existence of thermal operations can be ensured by usual majorization conditions, the existence
of GGPM requires additional conditions on $\rho_R$. For a general non-thermal $\rho_R$, Eq.~(\ref{eq:thermalOp}) may lead to a non-Gibbsian
$\rho_Q^\prime$ with coherences. We shall call the energy received by the thermometer system as effective heat if the state of the thermometer system is a Gibbsian. Such an operational definition is applicable to non-thermal sources, such as entangled atomic clusters as we consider here. After the interaction with the atomic cluster, a thermometer system can be found in a Gibbs state only for certain type of interactions and for special initial states of the atomic cluster. In the following section, we shall se that typical BE states can be perceived as artificial effective heat sources by a thermometer qubit under simple dipolar interaction. Initial coherences of BE states are translated only to the populations of the thermometer qubit and no coherences are injected in the energy basis of the qubit. 
For brevity we shall
simply call effective heat as heat. Operational single shot heat transfer scheme allows us to engineer effective temperatures of 
quantum systems using quantum resource states acting as artificial heat baths. It can be significant for engineering fast thermal
processes by using unitary interactions in compact quantum systems.

In the subsequent discussions we will show that BE states that we consider cannot induce any
coherence in $S_Q$ and leads to a Gibbsian $\rho_Q^\prime$  for a specific interaction between $S_R$ and $S_Q$.
We specifically consider random repeated applications of GGPM
in a micromaser scheme (cf.~Fig.~\ref{fig:schemes})
so that instead of small amount of heat transfer we can transfer larger amount of heat. Moreover we can determine the
maximum amount of heat ${\cal Q}$ that can be transferred under such a scheme. This approach determines the thermotropy of a set of copies of the
same BE states. Thermotropy depends on the form of $U_{RQ}$ and initial states $\rho_Q,\rho_R$.
Finding an optimum $U_{RQ}$ can be of fundamental interest
yet it is a non-trivial problem that we shall leave for future investigations. Instead
we follow a more practical route such that
$U_{RQ}$ will be fixed to a typical well-known (Tavis-Cummings or XX) interaction. Steady state temperature of $S_Q$ (micromaser cavity field) will be
used as a quantifier of ${\cal Q}$.

\section{Heat Extraction from Bound Entangled States}
\label{sec:heat}

\subsection{Single Interaction Scheme to Harvest Heat from BE States}
\label{sec:SS}
\begin{figure}[t!]
	\centering
	\includegraphics[width=8cm]{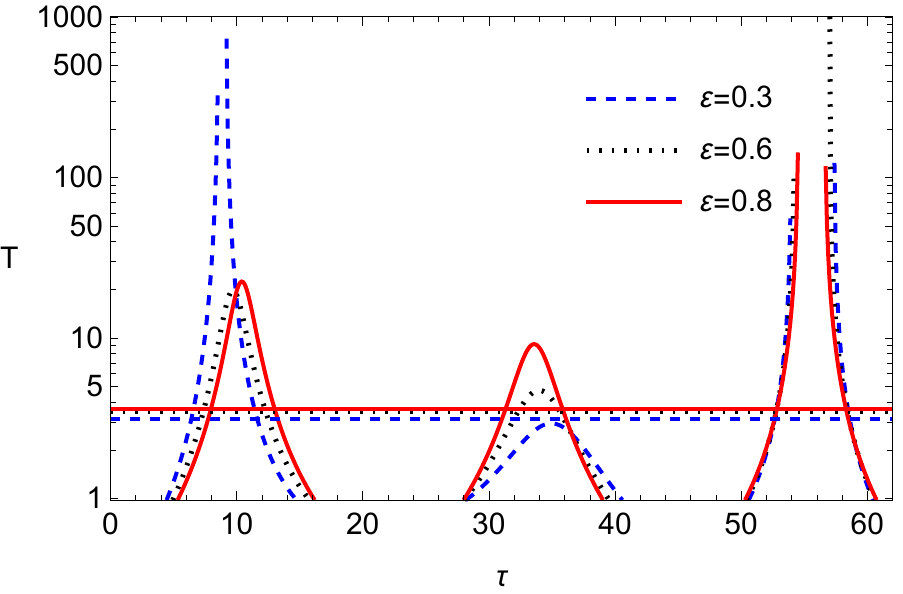}
	\caption{Effective temperatures (in the Log scale) of the target qubit for single-shot interaction model for initially different $\varepsilon$ values of the 
	FLS resource state. The initial temperature of the target qubit is zero. The horizontal lines represent the maximum temperature (thermotropy) values for different $\varepsilon$ 
	values in the repeated interaction scheme. $T/\omega$ is the unitless quantity. So the temperature has the unit of $[\omega]$ as $\hbar=k_B=1$. The dimensionless parameter, $\varepsilon$, is defined in the text.}
	\label{fig:fig_BE_T}
\end{figure}

As an example of heat extraction from a BE state using a single interaction, we consider FLS state of four qubits in Eq.~(\ref{eq:BES_matrix}). The 
harvester is taken to be a single qubit at the same frequency with the resource qubits for simplicity. The model hamiltonian is given by
$H_{\text{SS}} = H_\text{a} + H_\text{tq} + H_{\text{int}}$
where $(\hbar=1)$
\begin{eqnarray}
H_\text{a} &=& { \omega \over 2} \sum_{k=1}^N \sigma_k^z,
\label{eq:centralspin1}
\\
H_\text{tq} &=& \frac{\omega}{2}\sigma_0^z, \label{eq:centralspin2}
\end{eqnarray}
\begin{equation}
H_\text{int} =\Sigma_{i=1}^N g_i(\sigma_i^+\sigma_0^-+\sigma_i^-\sigma_0^+),\label{eq:centralspin}
\end{equation}
are the Hamiltonians of the atomic cluster, the target qubit (tq), and the interaction
between them, respectively. The interaction is assumed to be homogeneous case of the central spin model~\cite{Gaudin1976} with $g_i=g=0.1[\omega]$. 
The initial state of the total system is $\rho(0)=\rho_a(0)\otimes \rho_{\text{tq}}(0)$, where $\rho_a(0)$ and  $\rho_{\text{tq}}(0)$ are 
taken to be FLS BE state and a thermal state, respectively. Under the von Neumann-Liouvillian evolution, 
$\dot{\rho}=-i[H_{\text{SS}},\rho]$, the target qubit remains in a Gibbs thermal state and after an interaction time of $\tau$,
it evolves to $\rho_{\text{tq}}(\tau)$. Effective temperature $T(\tau)=1/\ln ( \rho_\text{tq}^{\text{grn}}(\tau)/\rho_\text{tq}^{\text{exc}}(\tau))$ of the target
qubit is determined from the populations of the excited and ground state levels of the qubit, given by $\rho_\text{tq}^{\text{exc}}(\tau) = \langle e |  \rho_\text{tq}(\tau) | e\rangle$ and 
$ \rho_\text{tq}^{\text{grn}}(\tau) = \langle g |  \rho_\text{tq}(\tau) | g\rangle$, respectively. Here, $ \rho_\text{tq}(\tau)=Tr_\text{a}\rho(\tau)$ is the reduced density matrix of the target qubit. The effective temperature is found at a higher temperature than
the initial temperature $T(0)$, as shown in Fig.~\ref{fig:fig_BE_T} for different $\epsilon$ values. We see that it is possible to
find interaction times for which BE states can outperform FE states to yield
higher temperatures, for example at short interaction times ($\tau < 10$). The figure
also shows the temperature that could be obtained under repeated short-time interactions with many 
copies of the resource qubits in FLS state (details of this scheme will be the subject of the next subsection). 
According to the operational definition of heat, transferred energy from the resource atomic cluster to the target qubit 
is perceived as heat. The energy transfer is reversible due to the interaction between the cluster and target qubit. Hence it can be optimized by the unitary evolution time $\tau$, as can be seen in Fig.~\ref{fig:fig_BE_T} where there are
sequences of intervals leading to higher temperatures than the ones obtained under repeated interactions. 
It is of practical significance that single-interaction time
can be tuned to certain values for which the heat transfer is more efficient than repeated interaction scheme. Moreover,
the control of relatively short interaction time ($\tau < 10$) allows for engineering a wide range of temperatures using BE states.

\begin{figure}[h!]
	\centering
	\begin{center}
	\subfigure[]{
	\label{fig:QvsTau}
	\includegraphics[width=0.4\textwidth]{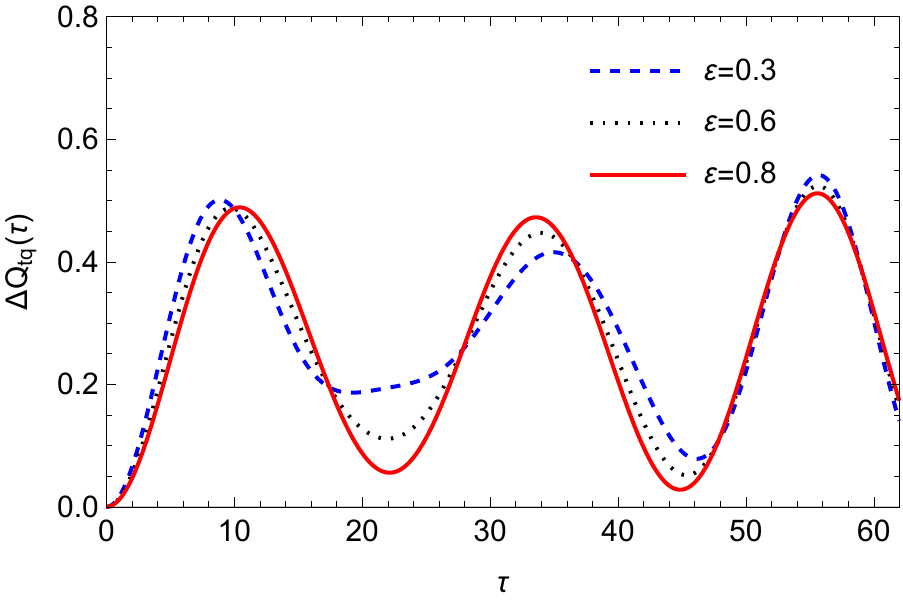}
	}
	\subfigure[]{
	\label{fig:SigmavsTau}
	\includegraphics[width=0.4\textwidth]{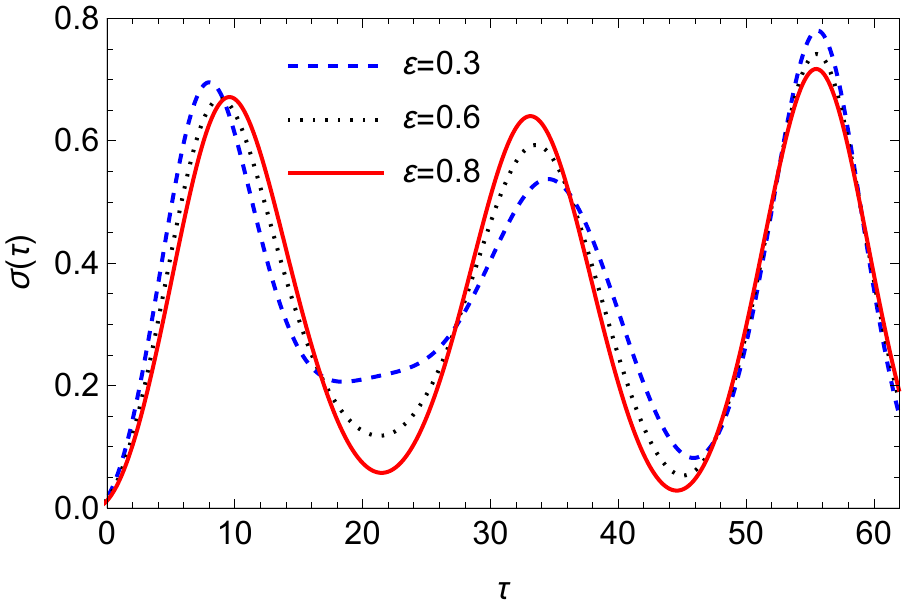}
	}
	\end{center}
	\caption{(a) Single-shot heat transfer $\Delta Q_{\text{tq}}$ to the target qubit depending 
	on the interaction time $\tau$ with a 4-atom resource state in an FLS
	state parametrized by $\varepsilon$.
	(b) Entropy production on the target 
	qubit depending 
	on the interaction time $\tau$ for different $\varepsilon$ values.}
\label{fig:fig_BE_Q}
\end{figure}

The heat transfer to the target qubit is determined 
by $\Delta Q_{\text{tq}}=\text{Tr}[H_{\text{tq}}\rho_{\text{tq}}(t)]-\text{Tr}[\rho_{\text{tq}}(0)H_{\text{tq}}]$. 
It is plotted in Fig.~\ref{fig:QvsTau}, for the cases of some FE and BE resources distinguished by different $\varepsilon$ parameter
in the FLS state. Behavior of the $\Delta Q_{\text{tq}}$ is qualitatively similar to the behavior of the $T(\tau)$ in Fig.~\ref{fig:fig_BE_T}.
While the target qubit temperature engineered by BE states can be infinitely high, the corresponding heat transfer is finite and comparable
to that of the FE states. 

As we use non-equilibrium resources and a single-shot operational heat transfer scheme, it is necessary to us to verify
the validity of second law of non-equilibrium thermodynamics. Our operational framework allows for a straightforward examination
of quantum Landauer principle for our finite size artificial bath and target qubit system. The change in the coherences of FLS BE state after an interaction time $\tau$ can be considered as an effective erasure, which contributes 
to the energy transfer and accordingly to the emergence of Landauer bound associated with the changes in the information content. Here we shall not examine information theoretic measures to quantify information content changes 
but directly evaluate the usual quantum Landauer bound given by the second law of the non-equilibrium thermodynamics~\cite{goold_nonequilibrium_2015} 
$\sigma\ge 0$, where $\sigma$ is the entropy
production 
$\sigma= \Delta S -\Delta Q/T$.
Here $\Delta S=S(\tau)-S(0)$ is the change in the von Neumann entropy $S=-\text{Tr}\rho\ln\rho$ of the target qubit. We remark that the Landauer bound is normally applied for a genuine heat bath at a genuine temperature. Here, however, we apply it to an effective artificial bath simulated by an atomic cluster; accordingly both the heat and the temperature are effective and operationally defined. Effective heat transfer from the atomic cluster can be calculated from the heat injection into the target (thermometer) system $\Delta Q$.
The effective bath temperature $T$ is taken as the final temperature of the target qubit.
This is justified by the energy conservation $[H_{\text{tq}}+H_a,H_{\text{int}}]=0$ in the central spin model Eqs.~(\ref{eq:centralspin1})-(\ref{eq:centralspin}). Another difference than the usual Landauer bound is that the temperature $T$ is not the temperature of the heat bath but an effective temperature we associate with the non-thermal atomic cluster. This temperature is taken to be the  same as the final temperature $T(\tau)$ of the target qubit at the end of its interaction with the resource qubits. Accordingly
the Landauer bound we investigate is in fact for the artifical bath that is engineered with the atomic cluster. We verify that artificial bath or the actual non-thermal system obeys the non-equilibrium second law in Fig.~\ref{fig:SigmavsTau} plotting $\sigma$ for 
different $\varepsilon$ values. The figure shows that $\sigma \geq 0$ at all interaction 
times and hence the second law holds for our non-equilibrium quantum information thermodynamical system.
The interaction times for which the entropy production deviates most from zero corresponds to the most efficient
heat transfer to the target qubit yielding maximum temperature increase. 

All of our observables, the temperature, the heat $\Delta Q$ and the
entropy production $\sigma$ show oscillations in time (Figs.~\ref{fig:fig_BE_T}-\ref{fig:fig_BE_Q}). The
non-decaying oscillatory behaviour can be traced back to the
integrability of our model; more specifically to the non-interacting
nature of the atomic cluster and the interaction set, e.g. XX-type,
between the target and the resource. The oscillatory behaviour can
also be regarded as a measure of the correlation between the target
and the resource; and hence when the heat quantifier approaches to
zero, both the target and the resource approximately return to their
initial uncorrelated (factorized) state. Deviations of $\sigma$ from
zero can be explained by finite mutual information
between the target and the resource, the finite free energy change of
the resource and the finite size corrections to the Landauer bound~\cite{reeb_improved_2014,esposito_entropy_2010,pezzutto_implications_2016}. The consistent behavior of $\sigma$ with $T(\tau)$ and $\Delta Q_{tq}$
in Figs.~\ref{fig:fig_BE_T}-\ref{fig:fig_BE_Q} verifies that the resource plays the role of an artificial heat bath faithfully within the laws of non-equilibrium
thermodynamics. Finally, we strictly distinguish the behavior that we
observe here in an artificial heat bath from a genuine heat bath which
is beyond the scope of our paper.

\subsection{Repeated Interaction scheme to harvest heat from BE states}
\label{sec:micromaser}
An analytical theory of heat exchange between a beam of quantum coherent atomic clusters and a micromaser cavity field has been developed originally for cluster sizes up to three qubits~\cite{dag_multiatom_2016,dag_temperature_2019}. Theoretical and numerical investigations for arbitrary size clusters have been presented subsequently~\cite{hardal_superradiant_2015}. Repeated interaction scheme to thermalize a single qubit using multi-qubit
clusters has also been discussed recently~\cite{manatuly_collectively_2019}.
A brief review of the micromaser case is given in the appendix~\ref{sec:app1} for notational clarity and to make the present discussion self-contained. 
In particular, it generalizes the earlier results to the cases of four qubits and two qutrits explicitly. 

The expression of the temperature of a micromaser cavity field pumped by a cluster of four atoms or two-qutrits  is determined in the appendix~\ref{sec:app1}  as
\begin{equation} 
\label{eq:repeatedIntTemp}
T(C,\delta) = { \hbar \omega_c \over k_B }
\left[ \ln \left( { R  + \delta + 2C  + 2 \kappa (\bar n_{\text{th}} +1) / \mu \over
R  - \delta + 2C  + 2 \kappa \bar n_{\text{th}}  / \mu} \right)
\right]^{-1},
\end{equation}
where $R = R_g + R_e=4$, $\delta = R_g - R_e$, $R_e= r_e-C$ and $R_g=r_g-C$ are introduced as shorthand notations.
Definitions of $r_e, r_g, C$ are given in  the appendix~\ref{sec:app1}. The crucial point to note is that
$\delta$ and $C$ depend only on populations and coherences of the cluster density matrix in the energy
basis, respectively. Those coherences that contribute to $C$ and change the cavity
field temperature without introducing any coherence to the field are called as 
heat exchange coherences (HECs)~\cite{dag_multiatom_2016}. Remarkably, all the BE states that we considered here 
possess only HECs. 
The expression Eq.~(\ref{eq:repeatedIntTemp}) has the same form with those obtained for the two and three 
qubit clusters used in the pump beam~\cite{dag_multiatom_2016,dag_temperature_2019}. 
Here we allow for the atomic clusters
to be subject to a generalized amplitude damping channel (GADC)~\cite{nielsen_quantum_2011} during their transfer to the micromaser cavity.
We remark that $R = R_g + R_e=4$ is the trace parameter, equal to number of atoms in a cluster, and it is invariant under GADC.
The steady state condition for which $T$ is well-defined is found to be $\delta+\kappa/\mu>0$, independent of coherences. This is
the same with the operation of the micromaser below threshold and amplification of incoherent (blackbody) 
radiation~\cite{scully_quantum_1967}.

The effect of GADC, which is a combination of amplitude damping channel and amplitude amplifying channel, on an atomic cluster during its transfer to the cavity
can be described in terms of Kraus operators $\mathbb{M}_{\text{GADC}}=(M_0, M_1, M_2,M_3)$ in the energy basis via Born-Markov approximation~\cite{louisell_1990,nielsen_quantum_2011}
\begin{eqnarray}\label{eq:gadc}
\mathbb{M}_{\text{GADC}} &=& \sqrt{\alpha} \left(\Ket{g}\Bra{g} + \sqrt{1-p_{\text{GADC}}} \Ket{e}\Bra{e} \right), \notag \\
&~&\sqrt{\beta} \left(\sqrt{1-p_{\text{GADC}}} \Ket{g}\Bra{g} + \Ket{e}\Bra{e} \right), \\ 
&~&\sqrt{\alpha p_{\text{GADC}}} \Ket{g}\Bra{e}, \sqrt{\beta p_{\text{GADC}}} \Ket{e}\Bra{g}, \notag
\end{eqnarray}
where $\alpha = \left(\bar{n}_{\text{th}}+1\right)/\left(2\bar{n}_{\text{th}}+1\right)$ and $\beta =\bar{n}_{\text{th}}/\left(2\bar{n}_{\text{th}}+1\right)$, which stand for ADC and AAC, respectively, and we assume that decoherence is identically applied to each atom separately. We describe the strength of the GADC at a given environment temperature as $p_{GADC} = 1-\exp[-\gamma t_{\text{tr}}(1+2\bar{n}_{th})/2]$~\cite{zhong_fisher_2013} given that $\gamma$ and $t_{tr}$ are the atomic damping rate and the transfer time of the atomic clusters to the cavity, respectively.

In Ref.~\cite{quan_quantum-classical_2006} microwave, optical, and superconducting resonator systems
are compared with each other and superconducting resonators are found to be the
most promising for effectively simulate micromaser system pumped with quantum
coherent atoms~\cite{liao_single-particle_2010}. Following these result, we use typical range of values for
our parameters in the simulations. We take $\kappa / \mu =1$ with the resonance frequency $\omega_c / 2 \pi = 10$ GHZ and
$T_{\mathrm{env}}\sim 160$ mK corresponds to $\bar n_{th} = 0.05$. For Cooper-pair box~\cite{wallraff_approaching_2005} or flux qubits ~\cite{stern_flux_2014}, we can take $\gamma / 2 \pi = 1$ MHZ.
In general, we will consider $t_{\text{tr}} = 50$ ns leading to $p_{\text{GADC}} \approx 0.15$, but we will also investigate the effect of $t_{\text{tr}}$ for a fixed $\gamma$.
Under these realistic conditions, below we will present the results on the heat extraction from four qubit Smolin and FLS BE states as well as
two qutrit Horedecki BE state.

Our approach of decoherence and heat extraction has limited applicability to specific classes of quantum states, we want to emphasize that in fact, this restriction is interesting. It is also closely related to the symmetry requirements of coherences in a quantum state in the energy basis that allows them to be exchanged as heat with a quantum thermometer system. It is pointed out that only certain coherences in a many-particle quantum state can be translated as heat into a probe system. These coherences are distributed in the blocks along the main diagonal of the density matrix in the energy basis~\cite{manatuly_collectively_2019}. The bound entangled states we considered here contain such coherences when expressed in the energy basis. 

Heat and work value of general two-particle and three-particle states have been stated, differences in Bell, W and GHZ type entanglement have been established~\cite{dag_temperature_2019, dag_multiatom_2016}, but we do not have a general form of bound entangled states; hence we have considered the most typical examples and parametric families of bound entangled states to make our results significant for related literature and experiments. Our operational definition heat is however general; its use in combination with decoherence for heat extraction from bound entangled states could be applicable for a broader class of many-particle entangled states depending on the progress on the revelations of relations among many-particle entanglement and many-particle coherences. Let us emphasize that an attractive feature of bound entangled states is their robustness to certain decay channels; they can emerge as a result of the decay of a free entangled state in such a channel. Accordingly, we may envision presented bound entangled states as natural, compact, robust quantum entangled fuels for quantum machines. 
\subsubsection{Smolin Bound Entangled State}
\label{sec:smolinHeat}

\begin{figure}[h]
\centering
\includegraphics[width=8cm]{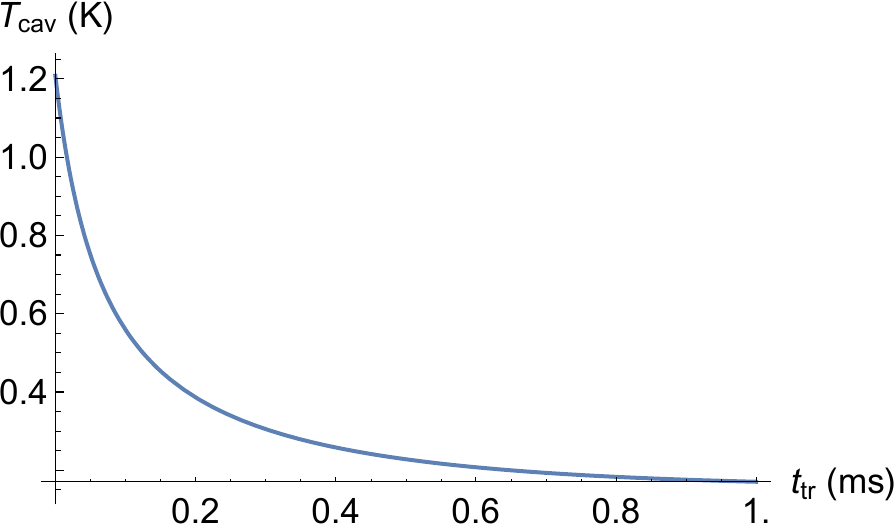}
\caption{Cavity temperature $T_\text{cav}$ (in Kelvin) achieved under pumping
	by four-atom clusters in Smolin state with respect to $t_{\text{tr}}$ (in ms),
	the transfer time of the atomic clusters to the cavity.}
\label{fig:fig_Smolin_ttr}
\end{figure}

The non-zero elements of the Smolin state are found as $a_{1,1}=a_{1,16}=a_{4,4}=a_{4,13}=a_{6,6}=a_{6,11}=a_{7,7}=a_{7,10}=a_{10,10}=a_{11,11}=a_{13,13}=a_{16,16}=1/8$ In the energy basis.  Accordingly we have $\delta=C=0$. Coherences in Smolin state are ineffective
in heat exchange under short time condition ($g\tau\ll 1$).
$\rho_\text{S}$ satisfies the conditions $\lambda=\xi=0$ so that no displacement or squeezing is
induced in the cavity field. While it cannot satisfy the threshold condition ($\delta+\kappa/\mu>1$) for the
heat exhange with a perfect cavity, this condition is satisfied with the help of cavity loss. It would act as a very hot bath while its effective temperature can
be controlled by exposing it to GADC before injecting into the cavity. After the GADC is applied for $t_{\text{tr}} = 50$ we find $\delta \approx 0. 58$. The cavity temperature is determined to be $T \approx 0.75$ K hence the copies of $\rho_S$ act as an effective
hot bath for the cavity field. The thermal gradient between the environment
of the cavity and the effective hot bath can be used to extract work efficiently
from the cavity
field using various quantum heat engine cycles which is an indirect route to
extract work from BE Smolin state  which is also an alternative to single shot cyclic action of a work source for ergotropy. In Fig.~\ref{fig:fig_Smolin_ttr}, we plot the dependence of cavity temperature on $t_{\text{tr}}$. As the exposure time of clusters to the environment $t_{\text{tr}}$, becomes longer (before they are injected into the cavity),  the cavity temperature approaches to the environment temperature of $160$ mK. This happens around $t_{\text{tr}}\sim 2$ ms for which $p_{\text{GADC}} \rightarrow 1$.

\subsubsection{FLS Bound Entangled State}
\label{sec:classHeat}

\begin{figure}[h]
\centering
\includegraphics[width=8.2cm]{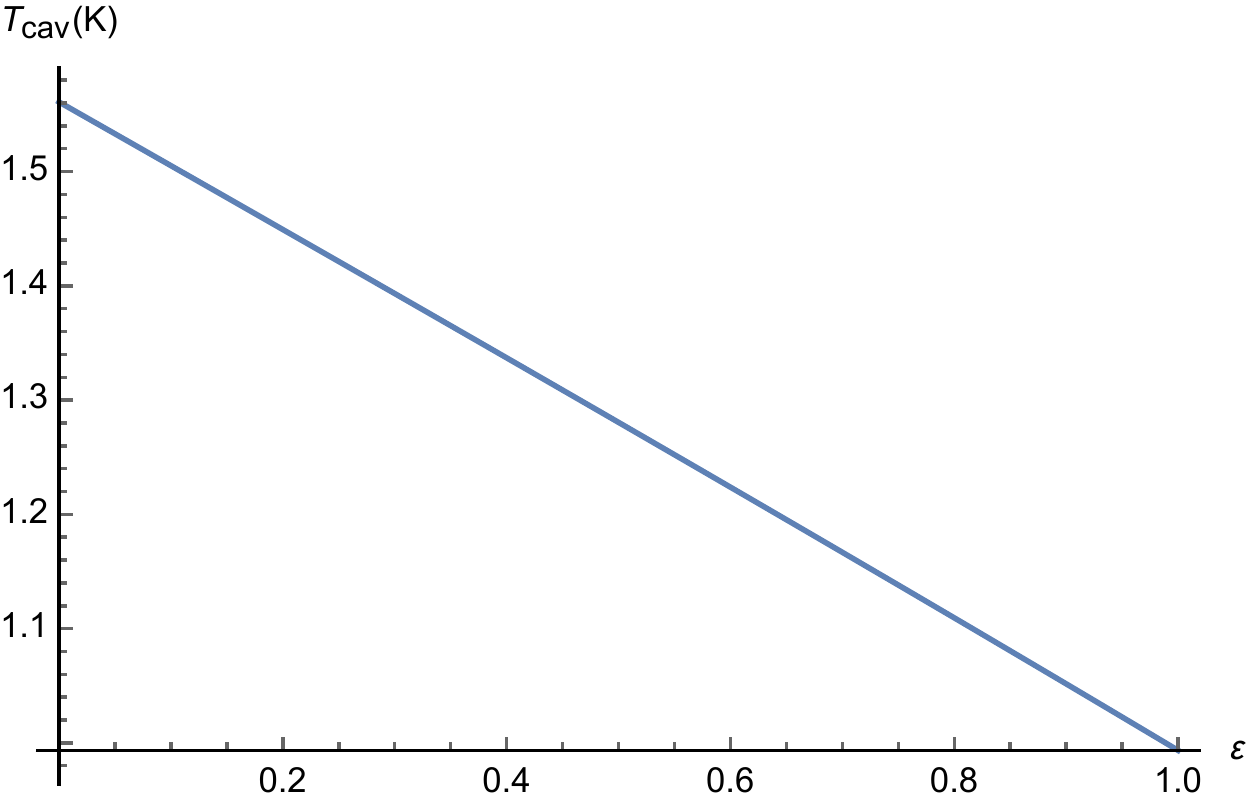}
\caption{Cavity temperature $T_\text{cav}$ (in Kelvin) achieved under pumping
	by four-atom clusters in a parametric entangled state with respect to the  $\varepsilon$, parameter (dimensionless) of the state, measuring distillable and nondistillable components of the entangled state. In the bound entangled region $0 \leq \varepsilon \leq 0.5$, the state leads to a higher cavity temperature than in the free entangled region $0.5 < \varepsilon \leq 1$.
	Exposure time of the atoms to the environment at $T_{\mathrm{env}}\sim 160$ mK before their injection into the cavity is $t_{\text{tr}} = 50$ ns. Here, $T_{\text cav}$ is unitless and scaled by $T_s = \hbar \omega / k_{\text B} = 0.48 K$ for $\omega_c /2\pi = 10 GHz$. The unitless parameter, $\varepsilon$, is defined in the text.
}
\label{fig:fig_Fei_varepsilon}
\end{figure}

\begin{figure}[h]
\centering
\includegraphics[width=8cm]{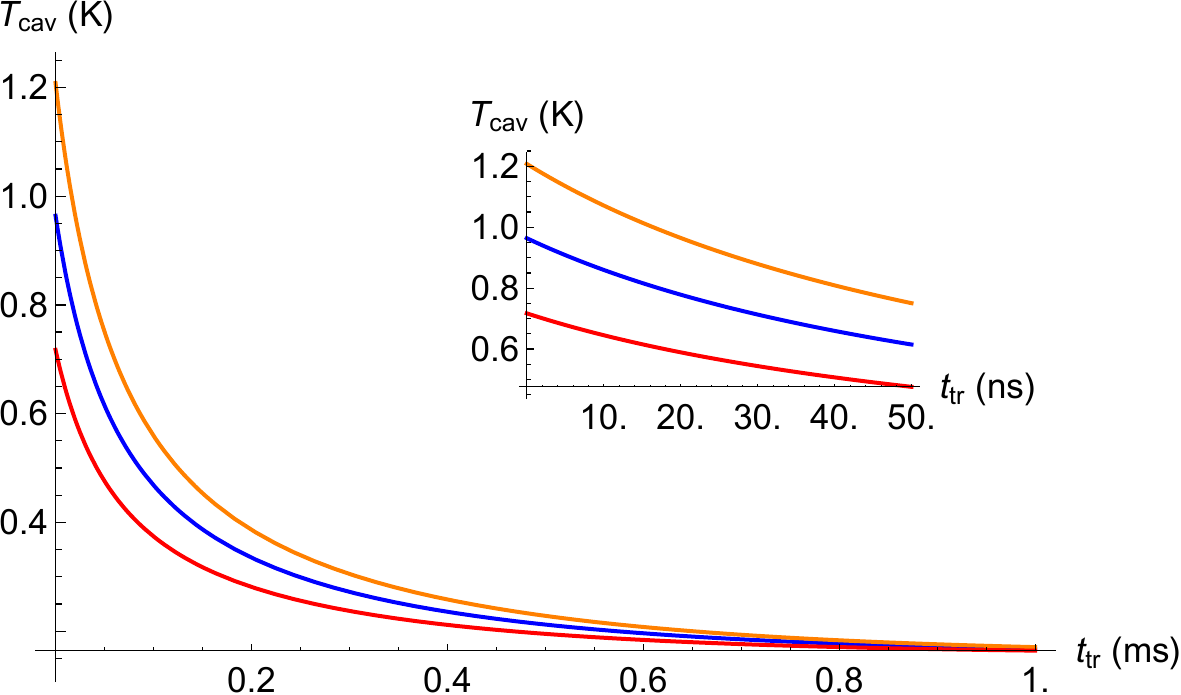}
\caption{Steady state temperature $T_\text{cav}$ (in Kelvin) of a micromaser cavity field pumped by four-atom clusters in FLS state $\rho_\varepsilon$ with respect to transit time $t_{\text{tr}}$ of the clusters through the cavity, for a bound entangled state at $\varepsilon = 0.5$ (blue middle curve) and a free entangled state at $\varepsilon = 1$ (red lower curve), and their completely dephased versions without any coherence (upper orange curve). The cavity temperature approaches to the temperature of the environment $T_\text{env}\sim 160$ mK as $t_{\text{tr}}\rightarrow 1$ ms. Inset shows the short time behavior up to $t_{\text{tr}}=50$ ns.}
\label{fig:fig_Fei_ttr}
\end{figure}

The non-zero elements of the class of bound entangled state are given in the excitation basis in Eq~(\ref{eq:BES_matrix}).
While this BE state has the same $\delta=0.58$ with the Smolin state, it has heat exchange coherences, that yields $C \approx - 0.85 \ \varepsilon$ after the action of GADC for an exposure time $t_{\text{tr}}=50$ ns. In addition to the populations, the negative coherence of this state contributes to
the control of the cavity temperature. For $t_{tr}=50$ ns an analytical expression
for the cavity temperature is found to be
\begin{equation}\label{eq:TcavFei}
T(\varepsilon) \approx \frac{\hbar\omega_c}{k_B} \left[ \ln \left(1 + {1.58 \over 1.76 - 0.85\varepsilon} \right) \right]^{-1},
\end{equation}
exhibiting a decreasing behavior with respect to $\varepsilon$, as shown in Fig.~\ref{fig:fig_Fei_varepsilon}. Monotonic and linear-like
behavior of $T_\text{cav}$ with $\varepsilon$ suggest that cavity temperature
can be used as a thermometer of such entangled states. Because the cavity temperature can clearly distinguish FE and BE from each other, in contrast to ergotropy
which is mirror-symmetric relative to $\varepsilon=0.5$
(cf.~Eq~(\ref{eq:ergotropyFei})). Higher cavity temperature is found in the BE region $0 \leq \varepsilon \leq 0.5$ compared to the free entangled region $0.5 < \varepsilon \leq 1$. Despite the negative
contribution of its coherences set of $\rho_\varepsilon$ act as an effective
hot bath for the cavity.
In order to see clearly the effect of the coherences in BE and FE-$\rho_\varepsilon$
states on the cavity temperature, let us consider  completely dephased versions of $\rho_\varepsilon$ at $\varepsilon=0.5$ and $\varepsilon=1$ before its transfer to the cavity, such that we eliminate all the the off-diagonal elements, but leave the diagonal ones untouched. The results
are plotted in Fig.~\ref{fig:fig_Fei_ttr}. 
The upper orange, the middle blue,
and the lowest red curves are for the cases of dephased, BE, and FE states, respectively. Negative coherences reduce the cavity temperature relative to the temperature contributed by the positive effect of populations. BE state remains superior to FE state at all exposure times to GADC channel.
As $t_{\text{tr}}$ increases, the strength of GADC increases so that the coherences decrease, which causes narrowing of the gap between the curves.

In addition to their completely dephased versions, FE and BE-$\rho_\varepsilon$ states
can be compared with other typical seprable states at the same energy. In Table~\ref{table:resultList}, we compare such states with each other and point to that neither of the FE statecan outperform the seperable states as their coherences
are negative. On the other hand, dependence of $T_\text{cav}$ on $\varepsilon$ allows us
to design a quantum heat engine cycle operating between two different heat baths whose temperatures engineered by two distinct sets of four-atom clusters with different distillability parameters $\varepsilon$. This can be used to harvest distillability as a resource.

\begin{table}[ht]
	\begin{center}
\small
\begin{tabular}{|c|c|c|c|}
  \hline
  State  & C & $\delta$ & T  \\ \hline
  $\rho_S$ & 0 & 0.58 & 0.75  \\ \hline
  $\rho_\varepsilon=0$     & 0 & 0.58 & 0.75   \\ \hline
  $\rho_\varepsilon=0.5$   & -0.42 & 0.58 & 0.61  \\ \hline
  $\rho_\varepsilon=1$     & -0.85 & 0.58 & 0.47  \\ \hline
  $|+\rangle \otimes|+\rangle \otimes |+\rangle \otimes |+\rangle$  & 2.52 & 0.58 & 1.53  \\ \hline
  $ ( \mathbb{1} \otimes \mathbb{1} \otimes \mathbb{1} \otimes \mathbb{1} )/16$  & 0 & 0.58 & 0.75  \\
  \hline
\end{tabular}
    \caption{Coherence $C$, population difference parameter $\delta$ and steady state cavity temperature $T$ achievable by some representative bound and free entangled as well as separable states. $\rho_S$ and $\rho_{\varepsilon=0},\rho_{\varepsilon=0.5},\rho_{\varepsilon=1} $ denote the Smolin state and  parametrized BE states, respectively.  For the separable states we take $\ket{+}=(\ket{g} +\ket{e})/\sqrt{2}$ and maximally mixed state $\mathbb{1}^{\otimes 4}/16$. All these 4-atom states are subject to GADC for a time $t_{\text{tr}}=50$ ns.}
    \label{table:resultList}
\end{center}
\end{table}

\subsubsection{Horodecki BE State of Atomic Qutrits }
\label{sec:HBET}

\begin{figure}[!h]
	\begin{center}
		\includegraphics[scale=0.65]{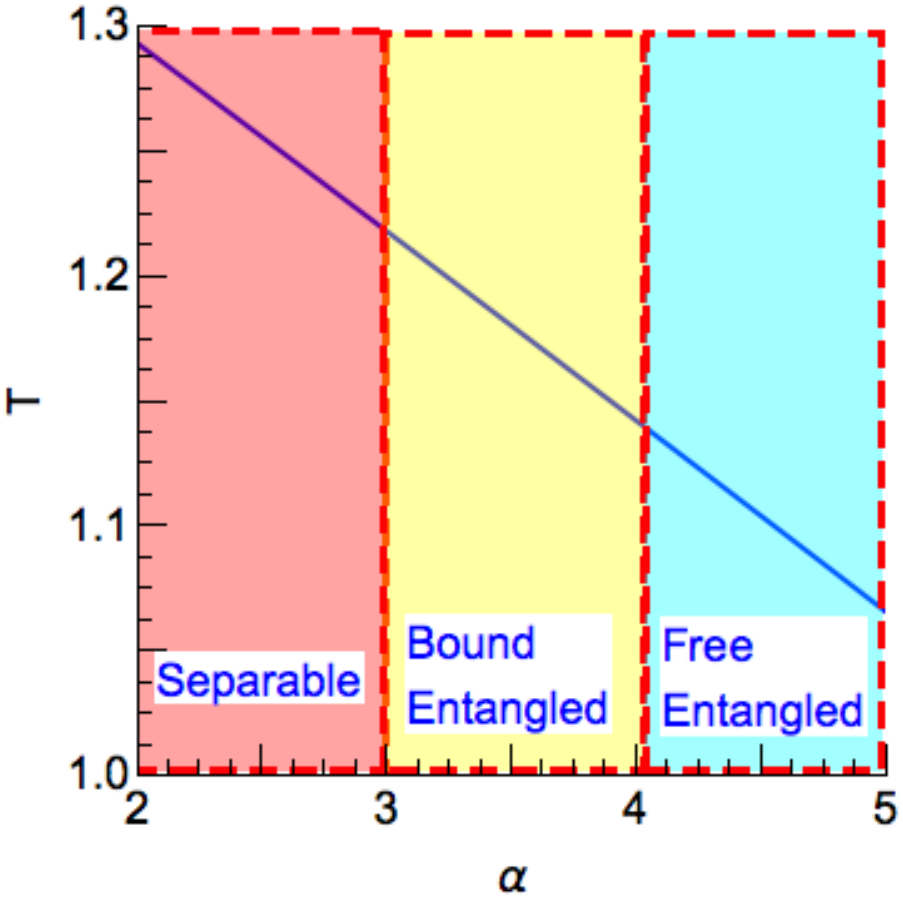}
		\caption{The effective temperature $T$ of the micromaser cavity field fueled with pair of 3-level atomic systems (qutrits), with respect to the $\alpha$ parameter determining whether the system is separable, bound entangled or free entangled.
			The micromaser temperature acts as an optical thermometer for measuring the entanglement of atomic qutrit pairs. 
			In the plot, $T$ is normalized with the cavity frequency and it is unitless, and $\hbar=k_B=1$.
			}
		\label{fig:freeBoundEntTermo}
	\end{center}
\end{figure}
If we use the Horodecki state~(\ref{eq:rho_alpha}), apart from the displacement coherences achieving $\lambda=4/21$, the excitation and de-excitation coefficients can be found as
\begin{eqnarray}
r_e=\frac{14-\alpha}{21},\quad r_d=\frac{28-\alpha}{21}
\end{eqnarray}
\noindent which imply that when this system is used as a fuel, it brings the cavity to a coherent thermal state.
The effective temperature $T$ for this system depends linearly on $\alpha$, as plotted in Fig.~\ref{fig:freeBoundEntTermo}.
In general, $\rho_\alpha$ transfers both heat and work to the cavity, and it is a suitable fuel for quantum thermo-mechanical machines. A lower (higher) temperature is achieved when fuel is free (bound) entangled. Therefore, although both have a worse performance than separable states, bound entanglement appears to be a better resource than free entanglement for heating purposes.
\section{Conclusions}
\label{sec:con}

We have investigated the work and heat value of typical four qubit and two qutrit BE states,
namely Smolin and FLS four qubit states and Horodecki two qutrit state. 

Maximal work extraction has been
quantified by an evaluation of the ergotropy. Qualitatively distinct behavior of ergotropy
with respect to distillability of 
entangled states have been found. For FLS state, FE and BE states have symmetric ergotropy values. Thus they share the same minimum ergotropy when distillable and
nondistillable entangled components of the state have the same weight.
For Horodecki state, FE regime exhibits stronger ergotropy than its BE component, though BE Smolin and FLS
states can have higher ergotropy than that of FE Horodecki state. We have also examined dynamics
of the ergotropy. Due to inevitable interactions with the environment, systems initially prepared in a FE state
may turn irreversibly into BE states, from which no entanglement can be distilled.
However, we showed that they can still possess significant ergotropy, indicating the significance of
nondistillable entanglement in the maximum work extraction in presence of decohering environments.

In order to explore heat extraction, we introduced
an operational definition and deduced the
conditions to identify the energy exchange with an
auxiliary (thermometer) system as heat. In addition to single-shot method, we also
considered a repeated interaction scheme to harvest heat. 
Defining maximum
heat transfer from a beam of BE states in repeated interactions with a micromaser cavity as
thermotropy, we characterized it in terms of steady state temperature of the cavity field. 
Smolin state is found to have no thermotropic value, yet it can still be used
as a fuel by exposing it to a decoherence channel before injecting into the cavity. 
FLS states have thermotropic value, yielding thermal cavity state with a proper
empirical temperature. 
Horodecki state delivers both work and
heat type energy in repeated interaction scheme yielding coherent thermal
steady state of the cavity field with an effective temperature. Both the proper and effective 
cavity temperatures for FLS and Horodecki states linearly decrease with the distilability, and hence allowing for thermometry of distilabillity of entanglement. Thermotropic value of 
BE states are found to be higher than FE states in FLS and Horodecki states. Accordingly,
In contrast to the typical choice of FE states for optimum work extraction, those BE states
are more favorable for heat extraction. Differences of distillability can be translated into effective
temperatures of the artificial heat baths simulated by BE atomic clusters so that distillability difference
in their entanglement can be harvested as work using a quantum heat engine. 

Moreover, we employed exact unitary dynamics to
examine the effect of interaction time in single shot route to pure heat extraction from FLS state. Superiority of 
BE FLS states over FE ones as effective thermal resources can be found at finite interaction 
times, too. BE states can outperform FE states to yield higher temperatures
at comparable level of heat transfer. It is found that 
one can engineer full positive temperature range using BE FLS states in contrast to FE
ones, which yields bounded temperatures at same interaction times. We verified that the energy
transfer from non-thermal states are within the Landauer bound if we assign the final temperature
of the target qubit as an effective temperature to the artificial bath simulated by entangled atomic cluster.
Single shot and repeated interaction
schemes of heat extraction have been compared. Single shot scheme can be used to obtain similar or higher temperatures
than the repeated interaction scheme and hence can be of practical significance to implement quantum thermalization
by unitary means. 

Our results can be significant to lead further avenues of practical applications
for bound entanglement and to illuminate fundamental relations among
irreversibility, distillability of entanglement, and energy processes in quantum information and
quantum thermodynamics. Our results can also be connected to the studies of entanglement and extractable energies, where local-operation and classical communication (LOCC) protocols are utilized. In such studies, either the work is extracted from a given classical resource, such as a thermal bath, with the help of entanglement, or it is harvested directly from quantum entanglement. For the former case, protocols using bipartite and tripartite entanglement have been proposed and a thermodynamical separability criterion, capable to distinguish W and GHZ states, is established~\cite{Viguie_2005}. For the latter case, an ancilla subsystem is typically considered, in addition to the resource  subsystem; they are not coupled but prepared initially in a joint quantum correlated state. Enhancement of ergotropy then becomes possible by performing local operations on the ancilla and using the feedback to improve work extraction characterized by so-called daemonic ergotropy~\cite{francica_daemonic_2017}. More recently, the difference between ergotropies of bipartite systems subject to global and local operations has been investigated rigorously, and such an ergotropic gap has been proposed as a witness for a limited class of bipartite entangled states~\cite{Alimuddin_2019}. Our approach, extracting work or heat from BE state shares the same objective with this latter case, but it is based upon global operations. For our four qubit or two-qutrit examples, we could also envision one of the qubits or qutrits in the system as an ancilla, then perform local measurements on it to extract work from the rest of the system. Regarding the heat extraction, advantages of using one of the qubits of a two-qubit state have been pointed out~\cite{Dillenschneider_2009}, then generalized to a multi-qubit system, where the role of ancilla-system entanglement on the enhancement of heat and work extraction has been revealed~\cite{turkpence_photonic_2017}. Rigorous examination of ergotropic gap for BE states can also be explored.
We hope our analysis here, focusing on global schemes of work and heat extraction from BE states, can inspire further studies along these directions.

\newpage
\begin{acknowledgments}
\"{O}. E. M. thanks to M. Paternostro for fruitful discussions. F. O. and \"{O}. E. M. acknowledge Isik University Scientific Research Fund, Grant No. BAP-15B103. F.O., A.T and \"{O}. E. M. acknowledge the support by TUBITAK, Grant No. 116F303 and by the EU-COST Action (CA15220).
\end{acknowledgments}
\section*{Appendix}
\appendix
\section{Smolin state eigenvectors which correspond to nonzero eigenvalues}
\label{sec:app0}
 In Sec~\ref{sec:smolinWork}, Smolin state is given as Eq.(~\ref{eq:Smolinstate}) and its nonzero eigenvalues are mentioned. Corresponding eigenvectors to nonzero eigenvalues related to Eq.(\ref{eq:W}) are  given below explicitly.
 \begin{eqnarray}
 |r_{1}\rangle =\frac{1}{\sqrt2}(|eegg\rangle+|ggee\rangle),\\
 |r_{2}\rangle =\frac{1}{\sqrt2}(|egeg\rangle+|gege\rangle),\\
 |r_{3}\rangle =\frac{1}{\sqrt2}(|eeee\rangle+|gggg\rangle),\\
|r_{4}\rangle = \frac{1}{\sqrt2}(|egee\rangle+|gegg\rangle).
   \end{eqnarray}
\section{Effective master equation for repeated interaction scheme}
\label{sec:app1}
The theory of repeated interactions of a target single mode cavity with atomic clusters follows some of the standard assumption of micromaser
scheme~\cite{meystre_elements_2007,scully_quantum_1997} (cf.~right panel of Fig.~\ref{fig:schemes}), where the
clusters are randomly injected into the cavity one at a time and their
time of transit through the cavity is faster than the photonic life
time but slower than the atom-photon interaction time; though we do not neglect neither the cavity loss nor the atomic decoherence. We assume the atoms will be subject to a generalized amplitude damping channel (GADC)~\cite{nielsen_quantum_2011} before they are injected into the cavity. The interaction of the atomic clusters with the cavity is described by the Tavis-Cummings model~\cite{tavis_exact_1968}
$H_{\text{TC}} = H_\text{a} + H_\text{c} + H_{\text{int}}$
where $H_\text{a}, H_\text{c}$ and $H_{\text{int}}$ are the the Hamiltonians of the atomic cluster, the cavity field and the interaction between them, which are given by
\begin{eqnarray}
H_\text{a} &=& { \hbar \omega_a \over 2} \sum_{k=1}^N \sigma_k^z,\\
H_\text{c} &=& \hbar \omega_c a^{\dag} a,
\end{eqnarray}
\begin{equation}
H_{\text{int}} = \hbar g \sum_{k=1}^N ( a \sigma_k^+ + a^{\dag} \sigma_k^-),
\end{equation}
respectively . Here $a$ and $a^{\dag}$ are the annihilation and creation operators for the cavity field, respectively, $\sigma_k^z, \sigma_k^+ , \sigma_k^- $ are the $z$, raising and lowering Pauli operators for the $k^{\text{th}}$ atom with $k=1,..., N$; and $g$ is the strength of photon-atom coupling, assumed to be spatially homogeneous. Atomic transition frequency $\omega_a$ is taken to be resonant with the cavity frequency $\omega_c$.

The unitary operator $U(\tau) = \text{exp}(-i H_{\text{int}} \tau)$ in the interaction picture for the system of the atomic cluster and the cavity can be obtained analytically up to second order in $g \tau$.
The evolution of the density operator of the cavity field obtained by tracing out the $j$'th atomic cluster injected in time $t_j$ is given as
a thermal operation in the form of Eq.~(\ref{eq:thermalOp}) such that
\begin{equation}
\rho(t_j + \tau) = Tr_a [ U(\tau) \rho_a \otimes \rho (t_j) U^{\dagger}(\tau) ] \equiv S(\tau) \rho(t_j),
\label{eq:TCMthermalOperation}
\end{equation}
where $S(\tau)$ is the superoperator that maps the cavity state from $\rho(t_j)$ to $\rho(t_j + \tau)$ at the end of the interaction time $\tau$, and $\rho_a$ is the initial density operator of the atomic clusters, which arrive randomly at a rate $p$. Probability to find a cluster inside
the cavity in a time interval $(t,t+\delta t)$ is $p\delta t$. Accordingly
the cavity field transformation can be expressed as a convex combination of
an identity and a partial energy exchange interaction such that
\begin{equation}
\rho(t + \delta t) = p \delta t S(\tau) \rho(t) + (1- p \delta t) \rho (t).
\end{equation}
The energy exchange component of this transformation is accomplished by the Tavis-Cummings model.
Its single qubit version, Jaynes-Cummings model~\cite{jaynes_comparison_1963}, has been studied
and recognized as an elementary form of a thermal operation which can be associated with a two-level
doubly stochastic matrix also known as T-transformation~\cite{lostaglio_elementary_2018}.
More generally, such
transformations can lead to equilibration between subsystems in terms of
distribution of resources, which is energetic value of quantum states in our case, and sometimes called as Robin-Hood transformations. Indeed we can write the corresponding transformation of atomic subsystem, similar to
Eq.~(\ref{eq:TCMthermalOperation}),
and see that cavity field acts
as a thermal bath to the atoms such that their coherences decrease (cf.~right panel of Fig.~\ref{fig:schemes}) and
their populations order as a thermal (Gibbs) distribution. Explicit conditions on getting a Gibbsian steady state for the cavity so that the
energy exchange can be identified as heat transfer will be discussed below.

In the limit of $\delta t\rightarrow 0$ we can find a master equation
of the form
\begin{equation}
\label{eq:rho_dot}
\dot{\rho} = p [S(\tau) -1] \rho(t),
\end{equation}
which can be written in the energy basis for four-atom clusters as
\begin{equation}\label{eq:mastereq1}
\dot{\rho}  = \left[ \sum^{16}_{i,j} a_{ij}   \sum^{16}_{n=1} U_{n i} (\tau) \rho(t) [ U_{n j} (\tau) ]^{\dagger} - \rho(t) \right],
\end{equation}
where $a_{i j}$ denote the elements of the density operator of $\rho_a$.
We remark that before injecting cluster into the cavity, we allow
atoms to be subject to noise from a generalized amplitude damping channel (GADC). Hence $a_{i j}$ denote
the elements after the action of GADC transformation as described in Eq.~(\ref{eq:gadc}).
We present the unitary propagator $U$ for four-atom clusters in the Appendix~\ref{sec:appB}.

In the presence of cavity loss, the master equation in Eq.~(\ref{eq:mastereq1}) can be written in the form
\begin{equation}
\dot{\rho} \approx - i [H_{\text{eff}}, \rho] + \mathbb{L}_s \rho + \mathbb{L}_c \rho +\mathbb{L} \rho,
\end{equation}
for sufficiently short $g\tau\ll 1$.

The effective Hamiltonian $H_{\text{eff}} = p g \tau (\lambda a^{\dagger} + \lambda^{*} a)$ describes as if a coherent drive applied to the cavity, and $\lambda$ denotes the sum of the coherences between the atomic states differing by one excitation.

The squeezing process denoted by the Lindbladian
$\mathbb{L}_s$ is given by $\mathbb{L}_s = \mu (\xi \mathbb{L}_s^e + \xi^* \mathbb{L}_s^d )$ with the effective coupling rate $\mu = p (g\tau)^2$. The squeezing excitation and de-excitation Lindbladians are
$\mathbb{L}_s^e = 2 a^{\dagger} \rho a^{\dagger} - a^{\dagger} a^{\dagger} \rho - \rho a^{\dagger} a^{\dagger} $, and
$\mathbb{L}_s^d = 2 a \rho a - a a \rho - \rho a a $.
$\xi$ is the sum of the coherences between states differing by two or three excitations.

The Lindbladian $\mathbb{L}$ is given in terms of the incoherent excitation Lindbladian
$\mathbb{L}_e = 2 a^{\dagger} \rho a - a a^{\dagger} \rho - \rho a a^{\dagger}$,
and de-excitation Lindbladian
$\mathbb{L}_d = 2 a \rho a^{\dagger} - a^{\dagger} a \rho - \rho a^{\dagger} a$, as
\begin{equation}
\mathbb{L} \rho = \mu\left(\frac{r_e}{2} \mathbb{L}_e\rho + \frac{r_g}{2} \mathbb{L}_d\rho \right).
\label{eq:LindbladianDefinition}
\end{equation}

The Lindbladian $\mathbb{L}_c$ describes the coupling of the cavity to the environment and is given by ~\cite{liao_single-particle_2010,quan_quantum-classical_2006,schaller_open_2014}
\begin{equation}
\mathbb{L}_{c}\rho = \frac{1}{2} \kappa \left(\bar{n}_{\text{th}} + 1\right) \mathbb{L}_d\rho + \frac{1}{2} \kappa \bar{n}_{\text{th}}
\mathbb{L}_e\rho,
\label{EqExtEnv}
\end{equation}
where $\kappa$ is the decay constant of cavity, $k_B$ is the Boltzmann constant, and $\bar{n}_{\text{th}}$ is the number of thermal photons in the environment
at temperature $T_\text{env}$,
\begin{equation}
\bar{n}_{\text{th}} = \frac{1}{e^{\hbar \omega_c/k_BT_\text{env}} -1}.
\end{equation}
In order to assign an empirical temperature
to the cavity field in the presence of cavity loss it is necessary to make sure $g\tau$ is sufficiently
small~\cite{liao_single-particle_2010}.

To ensure the energy exchange between the cavity field and the atomic cluster can be identified as heat first of all
the conditions $\lambda = \xi = 0$
has to be satisfied, otherwise the general solution of the master equation
would be of thermal squeezed state instead of Gibbsian.
Surprisingly BE states that we consider
satisfy these conditions.

When the first set of heat exchange conditions ($\lambda = \xi = 0$) are satisfied the master equation reduces to
\begin{eqnarray}\label{eq:repeatedIntME}
\dot\rho&=&\frac{\mu r_e+\kappa\bar n_{\text{th}}}{2}\mathbb{L}_e\rho+
\frac{\mu r_g+\kappa(\bar n_{\text{th}}+1)}{2}\mathbb{L}_d\rho.
\end{eqnarray}
The next condition to be satisfied is to operate the micromaser below
the maser threshold such that
$r_e<r_g+\kappa/\mu$,
for which steady state of the cavity field becomes a Gibbsian~\cite{scully_quantum_1967} with
an empirical temperature $T$
\begin{eqnarray}
\frac{r_e+\kappa\bar n_{\text{th}}/\mu}{r_g+\kappa(\bar n_{\text{th}}+1)/\mu}=e^{-\hbar\omega_c/k_BT},
\end{eqnarray}
whose solution for $T$ yields Eq.~(\ref{eq:repeatedIntTemp}).
While $T$ can be taken as a proper temperature for the cavity field, the atomic beam can be envisioned only as
an ``effective'' heat bath at $T$ which can be coherently engineered by heat exchange coherences (HECs) of the atomic cluster.

The parameters $\lambda, \xi$ for four qubit cluster are given as
\begin{eqnarray}
\lambda &=& a_{1,2}+a_{1,3}+a_{1,4}+a_{1,5}+  a_{2,6}+a_{2,9}+a_{2,10}+ a_{3,7}
+a_{3,9}+a_{3,11}+a_{4,8}+a_{4,10}\nonumber \\
&+&a_{4,11}+a_{5,6}+a_{5,7}+a_{5,8}+ a_{6,13}+a_{6,14}+ a_{7,13}+a_{7,15}+  a_{8,14}+a_{8,15} \nonumber \\
 &+&a_{9,12}+a_{9,13}+ a_{10,12}+a_{10,14}+  a_{11,12}+a_{11,15}+ a_{12,16}+a_{13,16}+a_{14,16}+a_{15,16}.\\
\xi &=& a_{1,6}+a_{1,7}+a_{1,8}+a_{1,9}+a_{1,10}+a_{1,11}+a_{2,12}+a_{2,13}+a_{2,14} \nonumber\\
&+&a_{3,12}+a_{3,13}+a_{3,15}+a_{4,12}+a_{4,14}+a_{4,15}+a_{5,13}+a_{5,14}+a_{5,15}\nonumber \\
&+&a_{6,16}+a_{7,16}+a_{8,16}+a_{9,16}+a_{10,16}+a_{11,16}.\\
r_e & = & 4 a_{11}      + 3 D_E + 2D_D + D_W + C \\
r_g & = & 4 a_{16,16} + 3 D_W + 2D_D + D_E + C,
\end{eqnarray}
where 
\begin{eqnarray}
D_E &=& \sum_{i=2}^{5} a_{ii}, \quad
D_D = \sum_{i=6}^{11} a_{ii}, \quad
D_W = \sum_{i=12}^{15} a_{ii}, 
\end{eqnarray}
and 
\begin{eqnarray}
C&=&C_E+ C_D + C_W,\\
C_E &=& \sideset{}{'}\sum_{i,j=2}^{5} a_{ij}, \quad
C_D = \sideset{}{'}\sum_{i,j=6}^{11} a_{ij} - \sideset{}{'}\sum_{i,j=6}^{11} b_{ij}, \quad
C_W = \sideset{}{'}\sum_{i,j=12}^{15} a_{ij},
\end{eqnarray}
Primed summations are constrained by $i \neq j$. $a_{ij}$ are the elements of density matrix of the qubit cluster in the energy basis, and 
$b_{ij}$ denote the anti-diagonal terms. Anti-diagonal elements of $\rho_a$ has no effect in the second order perturbation theory.

While for qutrit pair we have
\begin{eqnarray}
\lambda&=&\frac{1}{\sqrt{2}}\sum_{j=2,5}a_{1j}+\frac{1}{2\sqrt{2}}\sum_{i=2,5;j=6,9}a_{ij},\\
\xi&=&\frac{1}{2}\sum_{j=6,9}a_{1j},\\
r_e&=&\frac{1}{2}\left[ 4a_{11}+\sum_{i,j=2,3}a_{ij}+\sum_{i,j=4,5}a_{ij}+\sum_{i=4,5;j=2,3}(a_{ij}+a_{ji}) \right],\\
r_d&=&\frac{1}{2}\left[\sum_{i,j=6,9}a_{ij}+\sum_{i,j=2,3}a_{ij}+\sum_{i,j=4,5}a_{ij}+\sum_{i=4,5;j=2,3}(a_{ij}+a_{ji})\right].
\end{eqnarray}
For the qutrit pair, the  the matrix elements are in the energy basis ordered as $i,j=1,2,3,4,5,6,7,8,9 = ee, eu, eg, ue, uu, ug, ge, gu,gg$.
\section{Propagator of the four-qubit Tavis-Cummings model under short time approximation}
\label{sec:appB}
We list the matrix elements of the propagator in Eq.(\ref{eq:mastereq1}) in the short time approximation ($g\tau\ll 1$) in the energy basis as

\begin{equation}
\begin{split}
&U_{11} = 1-2 (g\tau)^2 \left(a^{\dagger} a + 1 \right), \\
&U_{21} = U_{31} = U_{51} = U_{91} = U_{42} = U_{62} = U_{10,2} = U_{43} \\
&\hspace{5.5mm}= U_{73} = U_{11,3} = U_{84} = U_{12,4} = U_{65} = U_{75} = U_{13,5} \\
&\hspace{5.5mm}= U_{86} = U_{14,6} = U_{87} = U_{15,7} = U_{16,8} = U_{10,9} = U_{11,9} \\
&\hspace{5.5mm}= U_{13,9} = U_{12,10} = U_{14,10} = U_{12,11} = U_{15,11} = U_{16,12}\\
&\hspace{5.5mm} = U_{14,13} = U_{15,13} = U_{16,14} = U_{16,15} = -i g\tau a^{\dagger} \\
&U_{12} = U_{13} = U_{15} = U_{19} = U_{24} = U_{26} = U_{2,10} = U_{34} \\
&\hspace{5.5mm}= U_{37} = U_{3,11} = U_{48} = U_{4,12} = U_{56} = U_{57} = U_{5,13} \\
&\hspace{5.5mm}= U_{68} = U_{6,14} = U_{78} = U_{7,15} = U_{8,16} = U_{9,10} = U_{9,11} \\
&\hspace{5.5mm}= U_{9,13} = U_{10,12} = U_{10,14} = U_{11,12} = U_{11,15} = U_{12,16}\\
&\hspace{5.5mm} = U_{13,14} = U_{13,15} = U_{14,16} = U_{15,16} = -i g\tau a \\
&U_{41} = U_{61} = U_{71} = U_{10,1} = U_{11,1} = U_{13,1} = U_{82} = U_{12,2} \\
&\hspace{5.5mm}= U_{14,2} = U_{83} = U_{12,3} = U_{15,3} = U_{16,4} = U_{84} = U_{14,4} \\
&\hspace{5.5mm}= U_{15,4} = U_{16,5} = U_{16,6} = U_{16,7} =  U_{12,9} = U_{14,9} = U_{15,9} \\
&\hspace{5.5mm}= U_{16,10} = U_{16,11} = U_{16,13} = - (g\tau)^2 \left(a^{\dagger} \right)^2 \\
&U_{14} = U_{16} = U_{17} = U_{1,10} = U_{1,11} = U_{1,13} = U_{28} = U_{2,12} \\
&\hspace{5.5mm}= U_{2,14} = U_{38} = U_{3,12} = U_{3,15} = U_{4,16} = U_{48} = U_{4,14} \\
&\hspace{5.5mm}= U_{4,15} = U_{5,16} = U_{6,16} = U_{7,16} =  U_{9,12} = U_{9,14} = U_{9,15} \\
&\hspace{5.5mm}= U_{10,16} = U_{11,16} = U_{13,16} = - (g\tau)^2 \left(a \right)^2\\
&U_{22} = U_{33} = U_{55} = U_{99} = 1- \frac{1}{2} (g\tau)^2 \left(4 a^{\dagger} a + 3 \right) \\
&U_{44} = U_{66} = U_{77} = U_{10,10} \\
&\hspace{5.5mm}= U_{11,11} = U_{13,13} = 1- \frac{1}{2} (g\tau)^2 \left(2 a^{\dagger} a + 1 \right) \\
&U_{88} = U_{12,12} = U_{14,14} = U_{15,15} = 1- \frac{1}{2} (g\tau)^2 \left(4 a^{\dagger} a + 1\right) \\
&U_{16,16} = 1- 2 (g\tau)^2 a^{\dagger} a \\
&U_{32} = U_{52} = U_{92} = U_{23} = U_{53} = U_{93} = U_{64} = U_{74} \\
&\hspace{5.5mm}= U_{10,4}  = U_{11,4} = U_{25} = U_{35} = U_{95} = U_{46} = U_{76} \\
&\hspace{5.5mm}= U_{10,6} = U_{13,6} = U_{47} = U_{67} = U_{11,7} = U_{13,7}= U_{12,8} \\
&\hspace{5.5mm}= U_{14,8}=U_{15,8} = U_{29}= U_{39} = U_{59} = U_{4,10} = U_{6,10} \\
&\hspace{5.5mm}= U_{11,10} = U_{13,10} = U_{4,11} = U_{7,11} = U_{10,11} = U_{13,11} \\
&\hspace{5.5mm} =  U_{8,12}= U_{14,12} = U_{15,12} = U_{6,13} = U_{7,13} = U_{10,13} \\
&\hspace{5.5mm} = U_{11,13}= U_{8,14} = U_{12,14} = U_{15,14} = U_{8,15}\\
&\hspace{5.5mm}= U_{12,15}= U_{14,15} = -\frac{1}{2} (g\tau)^2 \left(2 a^{\dagger} a + 1\right),
\end{split}
\end{equation}
and rest are all zero.



\end{document}